\documentclass[10pt,english]{IEEEtran}
\usepackage[T1]{fontenc}
\usepackage[latin9]{inputenc}
\usepackage{array}
\usepackage{verbatim}
\usepackage{amsmath}
\usepackage{graphicx}
\usepackage{amssymb}

\makeatletter

\providecommand{\tabularnewline}{\\}

\usepackage{dsfont}
\usepackage{array}
\usepackage{cite}
\usepackage{amsthm}
\usepackage{mathrsfs}
\makeatother
\usepackage{babel}
\makeatother

\usepackage{babel}

\begin{document}
\bibliographystyle{IEEEtran}

\title{An Upper Bound on Multi-hop Transmission Capacity with Dynamic Routing
Selection}

\author{Yuxin Chen and Jeffrey G. Andrews %
\thanks{Y. Chen is with the Department of Electrical Engineering and the Department
of Statistics, Stanford University, Stanford CA 94305, USA (email:
yxchen@stanford.edu). J. G. Andrews is with the Department of Electrical
and Computer Engineering, The University of Texas at Austin, Austin
TX 78712, USA (email: jandrews@mail.utexas.edu). The contact author
is J. G. Andrews. This research has been supported by the DARPA Information
Theory for Mobile Ad Hoc Networks (IT-MANET) program. It has been
presented in part at the IEEE International Symposium on Information
Theory 2010. Manuscript date: \today.%
}}
\maketitle
\begin{abstract}
This paper develops upper bounds on the end-to-end transmission capacity
of multi-hop wireless networks. Potential source-destination paths
are dynamically selected from a pool of randomly located relays, from
which a closed-form lower bound on the outage probability is derived
in terms of the expected number of potential paths. This is in turn
used to provide an upper bound on the number of successful transmissions
that can occur per unit area, which is known as the transmission capacity.
The upper bound results from assuming independence among the potential
paths, and can be viewed as the maximum diversity case. A useful aspect
of the upper bound is its simple form for an arbitrary-sized network,
which allows insights into how the number of hops and other network
parameters affect spatial throughput in the non-asymptotic regime.
The outage probability analysis is then extended to account for retransmissions
with a maximum number of allowed attempts. In contrast to prevailing
wisdom, we show that predetermined routing (such as nearest-neighbor)
is suboptimal, since more hops are not useful once the network is
interference-limited. Our results also make clear that randomness
in the location of relay sets and dynamically varying channel states
is helpful in obtaining higher aggregate throughput, and that dynamic
route selection should be used to exploit path diversity. \end{abstract}
\newtheorem{lemma}{\textbf{Lemma}}\newtheorem{theorem}{\textbf{Theorem}}\newtheorem{corollary}{\textbf{Corollary}}\newtheorem{remark}{\textbf{Remark}}

\section{Introduction}

In a distributed wireless network with random node locations, determining
the precise network capacity is a longstanding open problem that includes
many other simpler open problems as special cases \cite{ElGamalKim2011}.
Therefore, suboptimal analytical approaches that provide insights
into the achievable throughput and inform improved protocol design
are well-motivated, even if they fall short of strict upper bounds.
Multihop routing is generally considered necessary in large wireless
networks, both to ensure connectivity and to improve throughput, but
it is typically not well-optimized, nor are its performance limits
in general models known. In this paper we explore optimal multihop
strategies by considering dynamic path selection. Pre-determined routing
strategies such as nearest-neighbor routing, although they may perform
fairly well on average, are generally not optimal for a given network
state (which includes node positions and all the channels among them).
In fact, a randomly deployed set of potential relays with time-varying
fading channels may allow a large gain by providing more potential
routes. In this paper, we are interested in how the inherent randomness
in the network can be better harvested to improve the end-to-end success
probability and hence throughput over more static approaches.

We assume that node locations are a realization of homogeneous Poisson
process in order to investigate the throughput statistically. This
model provides a tractable way to characterize how the end-to-end
success probability and throughput varies over different multihop
routing strategies. We aim to study how multi-hop routing with the
assistance of a pool of randomly deployed relays impacts the throughput
scaling in a non-asymptotic regime, i.e. for networks of finite population
and area, with the goal of finding how much the average throughput
can be increased under quality of service (QoS) constraints. Considering
uncoordinated routing selection, we aim at determining the fundamental
limits for a general class of routing strategies instead of predetermined
selection. It can be expected that the diversity gain resulting from
the randomness and dynamic channels potentially provides significant
throughput improvement.

\subsection{Related Work and Motivation}

The best-known metric for studying end-to-end network capacity is
the transport capacity \cite{GupKum2000,JovVisKul04,XueXieKum05}.
This framework pioneered many notable studies on the limiting scaling
behavior of ad hoc networks with the number of nodes $n$ by showing
that the maximum transport capacity scales as $\Theta\left(\sqrt{n}\right)$
in arbitrary networks \cite{GupKum2000}. The feasibility of this
throughput scaling has also been shown in random networks by relaying
all information via crossing paths constructed through the network
\cite{FraDouTseTir2004}. Several other researchers have extended
this framework to more general operating regimes, e.g. \cite{OzgLevTse2007,OzgJohTseLev2008}.
Their findings have shown that nearest neighbor multihop routing is
order-optimal in the power-limited regime, while hopping across clusters
with distributed multiple-input and multiple-output (MIMO) communication
can achieve order-optimal throughput in bandwidth-limited and power-inefficient
regimes. However, most of these results are shown and proven for asymptotically
large networks, which may not accurately describe non-asymptotic conditions.
Moreover, scaling laws do not provide much information on how other
network parameters imposed by a specific transmission strategy affect
the throughput.

If node locations are modeled as a homogeneous Poisson point process
(HPPP), a number of results can be applied from stochastic geometry,
e.g. \cite{BacBlaMuh2006,BaccelliNOW}, in particular to compute outage
probability relative to a signal-to-interference-plus-noise ratio
(SINR) threshold. These expressions can be inverted to give the maximum
transmit intensity at a specified outage probability, which yields
the transmission capacity of the network \cite{WebYanAnd2005}. This
framework provides the maximum number of successful transmission the
network can support while simultaneously meeting a network-wide QoS
requirement. This framework allows closed-form expressions of achievable
throughput to be derived in non-asymptotic regimes, which are useful
in examining how various communication technique, channel models,
and design parameters affect the aggregate throughput, e.g. \cite{WebAndJin2007,WebAndYan2007,LiuAnd2011,BloJin09,HunAndWeb2008,VazHea09,YinGaoLiuCui09,HuaLauChen09};
see \cite{WebAndJin2008} for a summary. While the transmission capacity
can often be expressed in closed-form without resorting to asymptotics,
it is a single-hop or {}``snapshot'' metric. Recent work \cite{KouAnd2009,Ganti09isit}
began to investigate the throughput scaling with two-hop opportunistic
relay selection under different channel gain distribution and relay
deployment. However, more general multi-hop capacity has not proven
tractable.

If several other strong assumptions are made, e.g. that all relays
are placed equi-distant on a straight line and all outages are independent,
then closed-form multi-hop transmission capacity can be derived \cite{AndHaeKouWebJin2009}.
Stamatiou \emph{et. al.} \cite{StaRosHaeMavZeiZor09} also investigated
multihop routing in a Poisson spatial model, whose focus is to characterize
the end-to-end delay and stability, again based on predetermined routes.
`Other recent works analyzing the throughput of multihop networks
using stochastic geometric tools include \cite{NarKayLat2010}, which
extended \cite{AndHaeKouWebJin2009} to non-slotted ALOHA and \cite{Vaze2011},
which also adopted a similar framework to \cite{Vaze2011} to study
the throughput-delay-reliability tradeoff with an ARQ protocol, and
didn't require all hops to be equidistant. However, all of these used
predetermined routing selection. In fact, the outage of a predetermined
route does not preclude the possibility of successful communication
over other routes. Separately, multihop capacity has also been studied
in a line network without explicitly considering additional interference
\cite{SikLanHaeCosFuj06,OymSan2006}. This approach is helpful in
comparing the impact of additional hops in bandwidth and power-limited
networks, but fails to account for the interference inherent in a
large wireless network.

In addition, the above-mentioned diversity gain from dynamic relay
selection has been utilized for opportunistic routing \cite{BisMor2004}\cite{BacBlaMuh2008}
so any node that overhears packets can participate in forwarding.
Reference \cite{BacBlaMuh2008} appeared to be the first investigation
of the capacity improvement from opportunistic routing compared with
predetermined routing in a Poisson field. However, the performance
gain shown in \cite{BacBlaMuh2008} is based on simulation without
an exact mathematical derivation. Different random hop selection strategies
have also been studied and compared \cite{WebJinGanHae2008}\cite{Hae2005}
without giving tractable throughput bounds. Hence, characterizing
the available diversity gain is worth investigating. In this paper,
we will explicitly show that since a pool of randomly located relays
with varying channels provides more potential routes, more randomness
is preferable.

\subsection{Contributions and Organization}

Instead of predetermined routing, dynamic route selection from random
relay sets under varying channel states are investigated in this work.
The main contributions are summarized as follows.

(1) We provide a lower bound on the end-to-end outage probability
(Theorem \ref{thmGeneralOutBound}), which can be expressed as an
exponential function with respect to the expected number of potential
paths. This result implies that higher throughput can be achieved
when the correlation among the states of different hops is low and
hence randomness and opportunism is high.

(2) We further derive in closed-form the expected number of $\mathcal{S}-\mathcal{D}$
routes that can complete forwarding both for single transmission in
each hop and for two different retransmission strategies that are
subject to constraints on the number of allowed attempts, given in
Lemma \ref{thmAvgMultiRelay}. The basic idea is to map all relay
combinations to a higher dimensional space and focus on the level
set with respect to the success probability function.

(3) A closed-form upper bound on transmission capacity as a function
of outage constraint $\epsilon$ and the number of relays $m$ for
a general class of multi-hop routing strategies is given in Corollary
\ref{corollaryMultiRelay}, which follows from Lemma \ref{thmAvgMultiRelay}
and is the main technical result in the paper. These closed-form results
assume a general exponential form of success probability, which includes
most commonly used channel models as special cases, including path
loss, path loss with Rayleigh fading, and path loss with Nakagami
fading.

The above results show that in networks with uncoordinated routing,
an {}``ideal'' diversity gain arising from independent statistics
of different paths allows the throughput to exhibit near linear scaling
in the number of relays $m$ as long as the density of relay nodes
exceeds the threshold imposed by the outage constraint $\epsilon$.
This diversity gain requires strong {}``incoherence'' among different
paths, which would presumably degrade for large $m$ since longer
routes are more likely to be correlated or share common links. Unlike
the single-hop scenario where network throughput must decrease about
linearly as the output constraint $\epsilon$ is tightened, the multihop
capacity bound is less sensitive to $\epsilon$ especially for large
$m$. 

(4) Finally, we briefly show that all predetermined routing strategies
with no central coordination and without further information like
channel state information (CSI) may fail to outperform single hop
transmission in an interference-limited network because of the large
increase in interference. Hence, exploiting randomness is important
for multihopping to be viable in networks of finite size.

{}

The rest of the paper is organized as follows. In Section \ref{sec:Model},
we first define the end-to-end metric that quantifies the network-wide
throughput, and then state the key assumptions for the analysis, as
well as the channel models and their associated general form of per-hop
success probability. We then develop and derive lower bounds for end-to-end
outage probability for general scenarios in Section \ref{sec:OutageGeneral}.
Specifically, this provides a closed-form lower bound if the channel
model allows the per-hop success probability to be expressed in exponential
form, which is developed in Section \ref{sec:OutageExp}. This in
turn results in an upper bound for the multi-hop transmission capacity
in Section \ref{sec:TCBound}. Implications and interpretations of
the results are provided in Section \ref{sec:Discussion}.

\section{Models and Preliminaries}

\label{sec:Model}

\subsection{Models and Assumptions}

We assume that the locations of all sources are a realization of an
HPPP $\Xi_{t}$ of intensity $\lambda_{t}$, and a set of relays are
also randomly deployed in the plane with homogeneous Poisson distribution
independent of $\Xi_{t}$. We consider a fixed-portion model, i.e.,
the relay set is of spatial density $\frac{1-\gamma}{\gamma}\lambda_{t}$,
where $\gamma\in(0,1)$ is assumed to be a fixed constant. In other
words, if the locations of all wireless nodes are assumed to be an
HPPP $\Xi$ with intensity $\lambda$, then the set of active transmitters
is of intensity $\lambda_{t}=\gamma\lambda$. The destination node
is assumed to be a distance $R$ away from its associated source node
in a random direction, and is not a part of the HPPP. Suppose transmission
rate $b\approx\frac{1}{2}\log\left(1+\beta\right)$ is required for
successful transmission, where \textbf{$\beta$} is therefore the
required SINR. Denoting $\epsilon$ as the target outage probability
relative to $\beta$, the transmission capacity \cite{WebYanAnd2005}
in an uncoordinated single-hop setting is defined as \begin{equation}
T(\epsilon)=(1-\epsilon)\max_{\mathbb{P}(\text{SINR}<\beta)\leq\epsilon}\lambda_{t}\end{equation}
 which is the maximum expected throughput per unit area. Since $b$
is simply a constant function of $\beta$, we ignore it for simplicity.

Now suppose each session uses $k$ transmissions with the assistance
of the relay set. These $k$ attempts can be performed in an arbitrary
$k$ orthogonal slots, i.e., the unit time slot can be divided into
$k$ equal subslots and the source and relays take turns transmitting
in these subslots: only one transmitter per route is active at a time.
The contention density is still of density $\lambda_{t}$, but each
packet is transmitted $k$ times. Therefore, the multihop transmission
capacity metric should be modified to be \begin{equation}
T_{m}(\epsilon)=(1-\epsilon)\max_{\mathbb{P}(\text{SIR}<\beta)\leq\epsilon}\frac{\lambda_{t}}{k},\end{equation}
 since each hop requires a time slot, so the overall throughput must
be normalized by $k$. It should be noted that although one can {}``pipeline''
by simultaneously transmitting different packets on different hops,
this does not change the transmission capacity since the contention
density simply becomes $k$ times larger. Similar analysis can be
applied in quantifying the transmission capacity with this intra-route
spatial reuse but leads to the same result. When no retransmissions
are allowed, we have $k=m+1$ with $m$ relays; if we consider $M$
total attempts (including retransmissions) for any single session,
then $k=M$.

Slivnyak's theorem \cite{StoKenMec1996} states that an entire homogeneous
network can be characterized by a typical single transmission. Conditioning
on a typical pair, the spatial point process is still homogeneous
with the same statistics. Suppose that all transmitters employ equal
amounts of power, and the network is interference-limited, i.e. noise
power is negligible compared to interference power. Relays can be
selected from all nodes in the feasible region. In this paper, we
consider the effects of both path loss and fading. For point-to-point
transmission from node $i$ to node $j$ at a distance $r_{ij}$,
the requirement for successful reception in this hop is expressed
in terms of signal-to-interference ratio (SIR) constraint as \begin{equation}
\text{SIR}_{ij}=\frac{\left\Vert h_{ij}\right\Vert ^{2}r_{ij}^{-\alpha}}{\sum_{k\neq i}\left\Vert h_{kj}\right\Vert ^{2}r_{kj}^{-\alpha}}\geq\beta,\end{equation}
 where $\alpha$ denotes the path loss exponent, and $h_{ij}$ the
fading factor experienced by the path from $i$ to $j$. Distinct
links are assumed to experience i.i.d. fading, which is typically
reasonable.

\begin{table}
\caption{Summary of Notation and Parameters}

\centering{}\begin{tabular}{>{\raggedright}p{0.7in}>{\raggedright}p{2.4in}}
$\alpha$  & path loss exponent\tabularnewline
$\beta$  & SINR requirement for successful reception per hop\tabularnewline
$\epsilon$  & outage probability constraint\tabularnewline
$R$  & $\mathcal{S}-\mathcal{D}$ transmit distance\tabularnewline
$\lambda$  & contention density of all potential transmitters, intensity of $\Xi$\tabularnewline
$\lambda_{t}$  & contention density of active transmitters in any subslot, intensity
of $\Xi_{t}$\tabularnewline
$\gamma$  & the portion of active transmitters, $\lambda_{t}=\lambda\gamma$\tabularnewline
$r_{ij}$  & Euclidean distance from node $i$ to node $j$\tabularnewline
$K,$ $G$  & general parameters in exponential form single-hop success probability,
$g_{0}\left(r_{ij},\lambda_{t}\right)=G\exp\left(-\lambda_{t}Kr_{ij}^{2}\right)$\tabularnewline
$p_{\text{out}}^{(m)}$  & end-to-end outage probability employing $m$ relays without retransmissions\tabularnewline
$Z_{m}$  & the corresponding location vector in $\mathscr{R}^{2m}$ \tabularnewline
$g_{0}\left(r_{ij},\lambda_{t}\right)$  & single hop success probability \tabularnewline
$g_{m}\left(Z_{m},\lambda_{t}\right)$  & the end-to-end ($m+1$ hops) success probability with the relay set
$Z_{m}$\tabularnewline
$K,$ $G$  & general parameters in exponential form single-hop success probability,
$g_{0}\left(r_{ij},\lambda_{t}\right)=G\exp\left(-\lambda_{t}Kr_{ij}^{2}\right)$\tabularnewline
$v_{2m}\left(B\right)$  & Lebesgue measure of $B$\tabularnewline
$d_{m}(Z_{m})$  & sum of squared distance of $m+1$ hops with the relay set $Z_{m}$\tabularnewline
$D_{m}$  & distance constraint, $d_{m}\left(Z_{m}\right)\leq D_{m}$ \tabularnewline
$T_{m}\left(\epsilon\right)$ & $m+1$ hop transmission capacity\tabularnewline
$\Lambda$  & $\Lambda:={\lambda\gamma}K$\tabularnewline
$\kappa$  & $\kappa:=G\pi(1-\gamma)/\gamma K$\tabularnewline
\end{tabular}%
\end{table}

\subsection{Per-Hop Success Probability}

A Poisson node distribution often results in an exact or approximate
exponential form for per-hop successful probability. That is, given
that the packet is transmitted from node $i$ to next hop receiver
$j$ over distance $r_{ij}$ and contention density $\lambda_{t}$,
the probability that the received SIR stays above the target $\beta$
can be expressed as \begin{align}
\mathbb{P}\left(\text{SINR}_{ij}>\beta\right)\overset{\Delta}{=}g_{0}(r_{ij},\lambda_{t})=G\exp\left(-\lambda_{t}Kr_{ij}^{2}\right)\text{,}\label{eqnExpForm}\end{align}
 where $G$ and $K$ depend on the specific channel models and are
independent of $r_{ij}$ and $\lambda_{t}$. This holds for several
commonly used channel models, including Rayleigh fading with path
loss, Nakagami fading with path loss, and path loss without fading,
as we briefly show in this subsection before using the general form
in the remainder of this paper.

\subsubsection{Rayleigh fading}

Baccelli et. al. \cite{BacBlaMuh2006} showed under Rayleigh fading
that: \begin{align}
g_{0}(r_{ij},\lambda_{t})=\exp\left(-\lambda_{t}r_{ij}^{2}\beta^{2/\alpha}C(\alpha)\right)\text{,}\end{align}
 where $C(\alpha)=2\pi\Gamma(\frac{2}{\alpha})\Gamma(1-\frac{2}{\alpha})/\alpha$
with $\Gamma(z)=\int_{0}^{\infty}t^{z-1}\exp(-t)\mathrm{d}t$ being
the Gamma function. Hence, the coefficients under Rayleigh fading
can be given as \begin{align}
K_{\text{RF}}~=~\beta^{\frac{2}{\alpha}}C(\alpha);\quad G_{\text{RF}}~=~1.\end{align}

\subsubsection{Nakagami fading}

Nakagami fading is a more general fading distribution, whose power
distribution can be expressed in terms of fading parameter $m_{0}\geq~0.5$
as \begin{equation}
f_{Z}(z)=\frac{m_{0}^{m_{0}}z^{m_{0}-1}}{\Gamma(m_{0})}\exp(-m_{0}z).\end{equation}
 Recent work \cite{HunAndWeb2008} has suggested a way to study the
outage probability by looking at the Laplacian transform. In both
low-outage and high-outage regimes, the success probability under
Nakagami fading can be expressed as an exponential function with the
following coefficients\begin{align*}
\text{low outage regime: } & K_{\text{NF}}=\Omega_{m_{0}}\beta^{\frac{2}{\alpha}},~G_{\text{NF}}=1,\\
\text{high outage regime: } & K_{\text{NF}}=\Omega_{m_{0}}\beta^{\frac{2}{\alpha}},\\
 & G=1+\sum_{k=1}^{m_{0}-1}\sum_{l=1}^{k}\frac{l!}{k!}\left(\frac{-2}{\alpha}\right)^{l}\Upsilon_{k,l}\end{align*}
 Both the derivation of these coefficients and the definition of $\Omega_{m_{0}}$
and $\Upsilon_{k,l}$ can be found in Appendix \ref{AppendixNaka}.
These two regimes are typical in practical systems.

\subsubsection{Path Loss Model (non-fading)}

The exact closed-form formula of the success probability with only
path loss is unknown. One approach is letting $m_{0}\rightarrow\infty$
for Nakagami fading, which converges to a path-loss-only model. However,
lower and upper bounds follow an exponential form. For example, partitioning
the set of interferers into dominating and non-dominating nodes, an
upper bound can be obtained as $g_{0}^{\text{ub}}(r_{ij},\lambda_{t})=\exp(-\lambda_{t}\pi\beta^{\frac{2}{\alpha}}r_{ij}^{2})$.
Reference \cite{WebYanAnd2005} has also shown an upper bound on the
transmission capacity that is $\frac{\alpha}{\alpha-1}$ times the
lower bound for small $\epsilon$, and has illustrated by simulation
the tightness of these bounds. This suggests that there exists some
constant $K_{\text{PL}}$ such that \begin{align}
g_{0}(r_{ij},\lambda_{t})=\exp\left(-\lambda_{t}K_{\text{PL}}r_{ij}^{2}\right) & ,\end{align}
 where \begin{align}
\pi\beta^{\frac{2}{\alpha}}\leq K_{\text{PL}}\leq\frac{\alpha}{\alpha-1}\pi\beta^{\frac{2}{\alpha}},\quad G_{\text{PL}}=1 & .\end{align}

\section{Main Results }

\label{sec:MainResults}

In this section, we first address the end-to-end outage probability
for general channel models, which builds a connection between outage
probability and expected number of potential paths. Next, if channel
models allows the per-hop success probability to be expressed in an
exponential form, closed-form lower bounds can be derived. This in
turn will provide a closed-form upper bound for multihop transmission
capacity.

\subsection{Outage Probability Analysis for General Per-hop Success Probability}

\label{sec:OutageGeneral} 
\begin{figure}
\centering \includegraphics[width=3in]{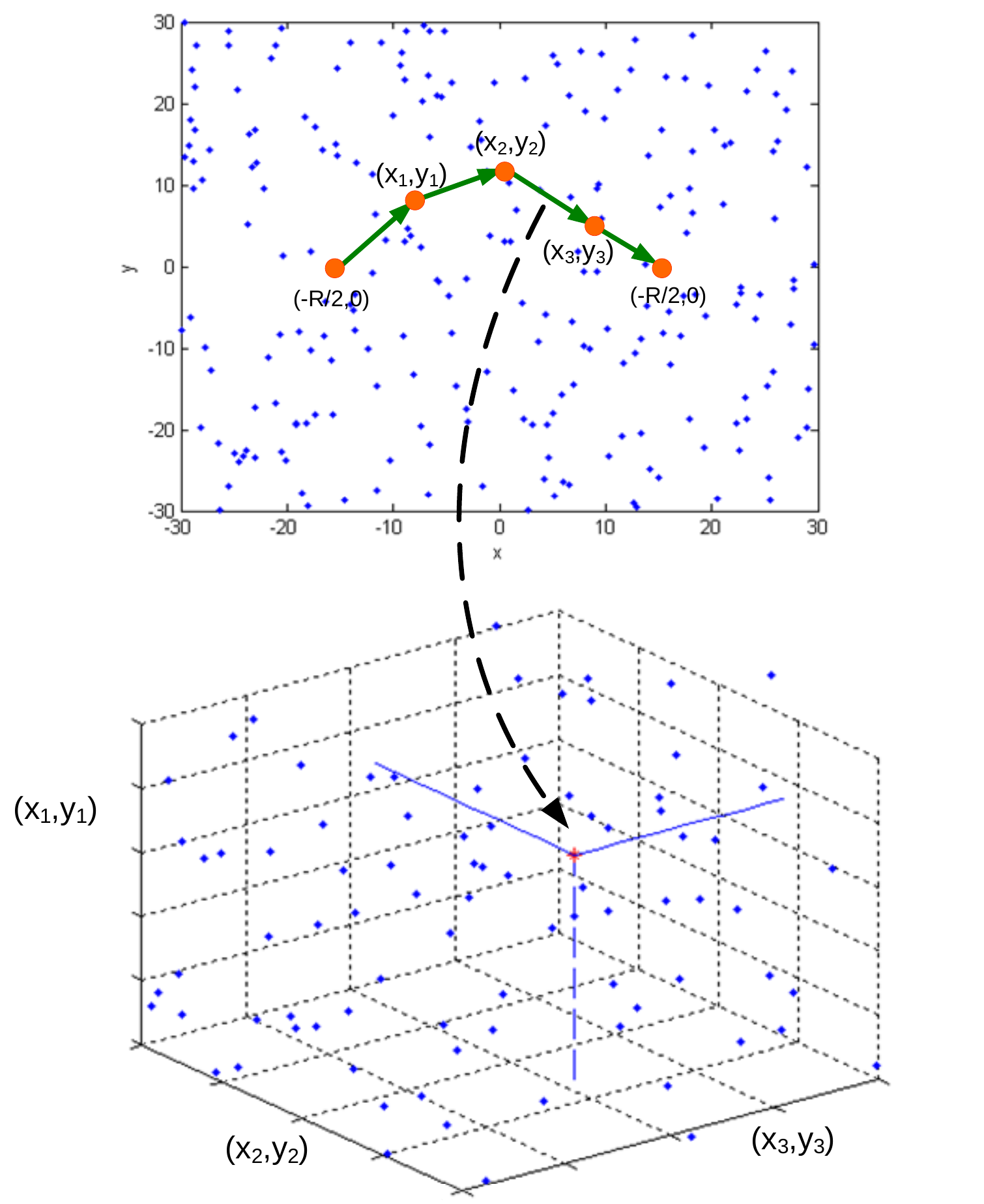}

\caption{In the left plot, the $\mathcal{S}-\mathcal{D}$ pair use $3$ relays
to assist the transmission, which can be matched to a point in high-dimensional
space as plotted in the right plot. In fact, a precise plot requires
drawing on $6$ dimensional space with each relay accounting for $2$
dimensions, but the right plot may help explain the mapping intuitively. }

\label{figHighDim} %
\end{figure}

Suppose that $m$ relays are employed by
a typical source-destination ($\mathcal{S}-\mathcal{D}$) pair. Since
all $\mathcal{S}-\mathcal{D}$ pairs are stochastically equivalent,
we can investigate the performance by looking at a typical $\mathcal{S}-\mathcal{D}$
pair. We will build a connection between the outage probability and
the expected number of relay sets that can connect the source and
destination. Suppose that there is a transmission pair with source
and destination located at $(-R/2,0)$ and $(R/2,0)$, respectively.
With the $i^{th}(1\leq i\leq m)$ relay located at $(x_{i},y_{i})$,
let $Z_{m}=(x_{1},y_{1},\dots,x_{m},y_{m})$ denote the location vector
of this specific relay set. From Slivnyak's theorem, conditional on
a typical transmission pair or finite number of nodes, the remaining
point process is still homogeneous Poisson process with the same spatial
density (we ignore a finite number of singular points here). Therefore,
all relay combinations form a homogeneous point process in a $2m$-dimensional
space $\mathcal{R}^{2m}$, as illustrated in Fig. \ref{figHighDim}.
The effective spatial density is $\tilde{\lambda}=(1-\gamma)\lambda$,
which characterizes the density of the pool of nodes that have not
been designated for specific transmissions. Assume that each relay
combination $Z_{m}$ can successfully assist in communication between
the $\mathcal{S}-\mathcal{D}$ pair with probability $g_{m}(Z_{m},{\lambda_{t}})$.
If we call a relay set that can successfully complete forwarding in
a given realiazation of the spatial process a \emph{potential relay
set}, then the expected number of potential relay sets in a hypercube
$B$, denoted by $N_{B}$, can be expressed as \begin{align}
 & \lim_{v_{2m}(B)\rightarrow0}\mathbb{E}(N_{B})\nonumber \\
= & \lim_{v_{2m}(B)\rightarrow0}\mathbb{E}\left(\sum_{Z_{m}\in B}\mathbb{I}\left(Z_{m}\text{ is a potential relay set}\right)\right)\nonumber \\
= & \text{ }\tilde{\lambda}^{m}g_{m}(Z_{m},\lambda_{t})v_{2m}(B)\text{,}\end{align}
 where $v_{2m}(B)$ denotes the Lebesgue measure of $B$ and ${\mathbb{I}}(\cdot)$
denotes indicator function. Let the random variable $N_{m}$ be the
number of relay sets that can complete forwarding using $m+1$ hops.
Then $\mathbb{E}(N_{m})$ is the expected number of different routes
that can successfully forward the packets for an $\mathcal{S}-\mathcal{D}$
pair. A larger $\mathbb{E}(N_{m})$ naturally leads to lower outage,
where we formalize in the following theorem.

\begin{theorem}\label{thmGeneralOutBound} Assume that all end-to-end
transmissions are achieved via $m+1$ hops with $m$ relays. The end-to-end
outage probability for any $\mathcal{S}-\mathcal{D}$ pair $p_{out}^{(m)}$
can be lower bounded as \begin{equation}
p_{out}^{(m)}\geq\exp(-\mathbb{E}(N_{m})).\label{eqnPoutMRelayGeneral}\end{equation}
 \end{theorem} \begin{IEEEproof} The key idea is to view the outage
event as the intersection of a set of \emph{decreasing events}. The
basic properties of decreasing events suggests a lower bound by treating
all these events as mutually independent. See Appendix \ref{sec:Proof-of-Thm-1}
for complete proof. \end{IEEEproof}

The lower bound can only be approached when the pool of potential
relay sets form a Poisson point process in a corresponding $2m$-dimensional
space (detailed in Appendix \ref{sec:Proof-of-Thm-1}), i.e. all potential
relay combinations independent from each other. In practice, route
selection for different source-destination pairs are not independent,
so the derived lower bound is not obtained by realistic routing strategies.
This result, however, indicates that low correlation among different
routes can reduce the outage probability in essence by enhancing diversity. 

We conjecture that this bound is tight and reasonable for small $m$
(e.g. the bound is exact for single relay case) but may be loose for
large $m$. This is because for a fixed pool of relays, the correlation
among different routes increases when the number of relays $m$ grows,
i.e. for large $m$, many routes are likely to share at least one
link. Also, the bound may become loose for low outage regime (e.g.
$\epsilon<10^{-3}$), because the outage bound is an exponential function
of $\mathbb{E}\left(N_{m}\right)$, where even constant factor difference
may result in an exponential gap. The advantage of this bound is to
allow us to see how the maximum allowable throughput scales with fixed
normal outage constraint in the non-asymptotic regime, which will
be shown later. In fact, some constant correction factor can also
be applied based on some simulation results without changing the scaling
of maximum contention density versus outage constraint. The expected
number of different routes $\mathbb{E}\left[N_{m}\right]$ plays an
important role, which will be calculated exactly in the following
subsection.

We caution, however, that the key assumption of independent path selection
cannot be achieved in practice. Potential routes are virtually coupled
and never independent from each other. For instance, failure of finding
a potential route for one session typically implies a lower success
probability for another session. But when uncoordinated routing selection
is employed for all $\mathcal{S}-\mathcal{D}$ pairs and when the
number of relays is reasonably small, the correlation is mitigated.
And this simplication assumption allows a reasonable closed-form bound
to be derived.

{}

\subsection{Outage Probability Analysis for Exponential-Form Per-hop Success
Probability}

\label{sec:OutageExp}

Now we begin to concentrate on the success probability of exponential
forms. When no retransmissions are adopted, if a specific route is
selected for packet delivering over $m$ relays with hop distances
$r_{1}$, $r_{2},\dots,$ $r_{m+1}$, respectively, the probability
for successful reception can be found as the product of each hop's
success probability: \begin{align}
 & g_{m}(r_{1},\cdots,r_{m+1},\lambda_{t})\\
\text{=} & \prod_{i=1}^{m+1}g_{0}(r_{i},\lambda_{t})=G^{m+1}\exp\left(-{\lambda_{t}}K\sum_{i=1}^{m+1}r_{i}^{2}\right)\end{align}
 Here, we assume independence among the success probability of each
hop, which requires sufficient diversity in the interferer locations
over time. Although this assumption is not valid in practice since
transmissions across subsequent hops are correlated, it allows us
to retain tractability and has been shown to be a reasonable approximation
in uncoordinated networks (as illustrated in \cite{AndHaeKouWebJin2009}).
The temporal correlation may also be mitigated through diversity techniques
like frequency hopping. 

Conditional on a typical transmission pair with source and destination
located at $(-R/2,0)$ and $(R/2,0)$, the spatial point process is
still a HPPP with the same statistics. In the $m$ relay case with
the $i^{th}(1\leq i\leq m)$ relay located at $(x_{i},y_{i})$, let
$Z_{m}=(x_{1},y_{1},\dots,x_{m},y_{m})$ denote the locations of the
specific relay set, then we can define the corresponding distance
statistics as \begin{align}
d_{m}(Z_{m})\overset{\Delta}{=} & (x_{1}+\frac{R}{2})^{2}+\sum_{i=1}^{m-1}(x_{i}-x_{i+1})^{2}+(x_{m}-\frac{R}{2})^{2}\nonumber \\
 & \quad+y_{1}^{2}+\sum_{i=1}^{m-1}(y_{i}-y_{i+1})^{2}+y_{m}^{2}\text{.}\end{align}
 This is the sum of squares of hop distances. Hence, the routing success
probability for a specific set of relays with location $Z_{m}$ can
be explicitly expressed as \begin{equation}
g_{m}(Z_{m},\lambda_{t})=G^{m+1}\exp(-\lambda_{t}Kd_{m}(Z_{m}))\text{.}\end{equation}
 In fact, an arbitrary set of relays will have positive probability
for successful forwarding. However, for those relay sets with large
$d_{m}(Z_{m})$, the communication process becomes extremely fragile
and difficult to maintain due to the low reception probability and
large distance. Practical protocols usually attempt to search potential
routes inside a locally finite area instead of from the infinite space
since the longer routes are very unlikely to be an efficient one.
In order to leave the analysis general, we impose a constraint $d_{m}(Z_{m})\leq D_{m}$
for the $m$ relay case, where $D_{m}\rightarrow\infty$ reverts to
the unconstrained distance case. We will later show that a reasonably
small constraint $D_{m}$ is sufficient to achieve an aggregate rate
arbitrarily close to the capacity upper bound.

Moreover, since only one transmitter is active at a time along the
entire multihop route, in each subslot, each node is used as a relay
by other source-destination pairs with probability $\gamma$. Therefore,
the pool of relays in each hop can be treated as the original point
process $\Xi$ with each point being deleted with probability $\gamma$.
Hence, the location of all relay sets in $\mathscr{R}^{2m}$ can be
viewed as a realization of a point process with effective spatial
density $\lambda^{m}(1-\gamma)^{m}$. This leads to the following
lemma.

\begin{lemma}\label{thmAvgMultiRelay}Define $\kappa=G\pi(1-\gamma)/\gamma K$
and $\Lambda={\lambda\gamma}K$. If all end-to-end transmissions are
achieved via $m+1$ hops with $m$ relays, with a constraint $d_{m}(Z_{m})\leq D_{m}~(Z_{m}\in\mathscr{R}^{2m})$,
the expected number of potential relay sets can be computed as \begin{align}
\mathbb{E}(N_{m}) & =\frac{G\kappa^{m}}{m+1}\left\{ \exp\left(-\frac{\Lambda R^{2}}{m+1}\right)-\exp\left(-\Lambda D_{m}\right)\cdot\right.\nonumber \\
 & \quad\quad\quad\left.\sum\limits _{0}^{m-1}\frac{1}{i!}\left(\Lambda\left(D_{m}-\frac{R^{2}}{m+1}\right)\right)^{i}\right\} .\end{align}
\end{lemma}

\begin{IEEEproof}The key point in the proof is that the isosurface
of $d_{m}(Z_{m})$ forms a high-dimensional elliptical surface, which
provides a tractable closed-form solution. See Appendix \ref{sec:Proof-of-Lemma-AvgMultiRelay}.\end{IEEEproof}

This result indicates that a larger $m$ typically provides more diversity,
because it provides more possible combinations of different relays,
and the dynamically changing channel states provide more opportunities
for us to find a potential route. A larger feasible range for route
selection $D_{m}$ also increases the expectation, but since the effect
of $D_{m}$ mainly exhibits as an exponentially vanishing term, it
can be expected that a fairly small range is enough to approach the
limits. Moreover, this analytic framework can be extended to account
for retransmissions in the following two scenarios. First, the best-effort
retransmission protocol requires that each hop adopts $k$ retransmissions
regardless of the results of each transmissions. The following lemma
provides more general results for best-effort protocols by allowing
each hop to adopt a different number of retransmissions. Second, instead
of specifying retransmissions for each hop, we bound the maximum number
of total allowed attempts to $M$. Define ${\bf 1}:=\left(1,\cdots,1\right)^{T}$.
The following lemma provides closed-form results for these two scenarios.

\begin{lemma}\label{lemmaRetran}

Assume that all end-to-end transmissions are achieved via $m+1$ hops
with $m$ relays. 

(1) In the best effort retransmission setting, if the $i^{\text{th}}(1\leq i\leq m+1)$
hop is retransmitted $k_{i}$ times, then the expected number of potential
relay sets can be given as \begin{align}
\mathbb{E}_{{\bf {k}}}(N_{m})~ & =~\sum_{{\bf n}:\text{ }{\bf 1}\preceq{\bf n}\preceq{\bf k}}\frac{\left(-1\right)^{m+1}\pi^{m}\left(1-\gamma\right)^{m}\left(-G\right)^{\sum_{j=1}^{m+1}n_{j}}}{\gamma^{m}K^{m}}\cdot\nonumber \\
 & \quad\quad\quad\quad\prod_{i=1}^{m+1}\left(\begin{array}{c}
k_{i}\\
n_{i}\end{array}\right)\frac{m\exp\left(-\frac{\Lambda R^{2}}{\sum_{j=1}^{m+1}1/k_{j}}\right)}{\left(\prod_{i=1}^{m+1}n_{i}\right)\left(\sum_{i=1}^{m+1}\frac{1}{n_{i}}\right)}.\label{eq:ExBestEffort}\end{align}

(2) If the $\mathcal{S}-\mathcal{D}$ transmission allows $M$ transmissions
in total without specifying the number of retransmissions for each
hop, then the expected number of potential relay sets can be given
as \begin{align}
\mathbb{E}_{M}(N_{m}) & =\sum_{\substack{{\bf {k}}^{T}\cdot{\bf {1}}\leq M\\
{{\bf {k}}\succeq{\bf {1}}}}
}(-1)^{{\bf {k}}^{T}\cdot{\bf {1}}}\sum_{\substack{{\bf {j}}^{T}\cdot{\bf {1}}\leq M\\
{\bf {j}}\succeq{\bf {k}}}
}\prod_{l=1}^{m+1}{j_{l}-1 \choose k_{l}-1}\mathbb{E}_{{\bf {k}}}(N_{m}),\end{align}
where $\mathbb{E}_{{\bf k}}\left(N_{m}\right)$ is given in (\ref{eq:ExBestEffort}).

\end{lemma}

\begin{IEEEproof} See Appendix \ref{sec:Proof-of-Lemma}. \end{IEEEproof}

Similarly, this lemma is derived by mapping all potential relay sets
onto a $2m$-dimensional space and investigating the isosurfaces in
that space. The above results on the expected number of potential
relay sets immediately yield the following corollary.

\begin{corollary}\label{thmMultiRelay} Assume that all end-to-end
transmissions are achieved via $m+1$ hops with $m$ relays.

(1) If only a single transmission is allowed for in each hop, then
the outage probability under a constraint $d_{m}(Z_{m}){\leq}D_{m}~(Z_{m}{\in}\mathscr{R}^{2m})$
can be computed as \begin{align}
p_{\text{out}}^{(m)} & \geq\exp\left\{ -\frac{G\kappa^{m}}{m+1}\left\{ \exp\left(-\frac{\Lambda R^{2}}{m+1}\right)-\right.\right.\nonumber \\
 & \quad\left.\left.\exp\left(-\Lambda D_{m}\right)\sum\limits _{0}^{m-1}\frac{1}{i!}\left(\Lambda\left(D_{m}-\frac{R^{2}}{m+1}\right)\right)^{i}\right\} \right\} .\label{eqnPoutMRelay}\end{align}

(2) If each hop adopts $k_{i}(1\leq i\leq m+1)$ retransmissions,
the outage probability $q_{{\bf {k}}}$ can be computed as \begin{align}
q_{{\bf {k}}}\geq~\exp\left(-\sum_{{\bf n}:\text{ }{\bf 1}\preceq{\bf n}\preceq{\bf k}}\frac{\left(-1\right)^{m+1}\pi^{m}\left(1-\gamma\right)^{m}\left(-G\right)^{\sum_{j=1}^{m+1}n_{j}}}{\gamma^{m}K^{m}}\cdot\right.\nonumber \\
\prod_{i=1}^{m+1}\left(\begin{array}{c}
k_{i}\\
n_{i}\end{array}\right)\frac{m\exp\left(-\frac{\Lambda R^{2}}{\sum_{j=1}^{m+1}1/k_{j}}\right)}{\left(\prod_{i=1}^{m+1}n_{i}\right)\left(\sum_{i=1}^{m+1}\frac{1}{n_{i}}\right)}.\quad\quad\end{align}

(3) If the transmission adopts $M$ retransmission in total, the outage
probability $q_{M}$ can be given as \begin{align}
q_{M}\geq~\exp\left(-\sum_{\substack{{\bf {k}}^{T}\cdot{\bf {1}}\leq M\\
{{\bf {k}}\succeq{\bf {1}}}}
}(-1)^{{\bf {k}}^{T}\cdot{\bf {1}}}\cdot\quad\quad\quad\quad\right.\nonumber \\
\sum_{\substack{{\bf {j}}^{T}\cdot{\bf {1}}\leq M\\
{\bf {j}}\succeq{\bf {k}}}
}\prod_{l=1}^{m+1}{j_{l}-1 \choose k_{l}-1}\mathbb{E}_{{\bf k}}\left(N_{m}\right),\end{align}
where $\mathbb{E}_{{\bf k}}\left(N_{m}\right)$ is given in (\ref{eq:ExBestEffort}).
\end{corollary}

This corollary provides closed-form lower bounds on the end-to-end
outage probability. For sufficiently large $D_{m}$ in the non-retransmission
case, the lower bound reduces to: \begin{equation}
p_{\text{out}}^{(m)}\geq\exp\left\{ -\frac{G^{m+1}{\pi}^{m}(1-\gamma)^{m}}{{\gamma}^{m}K^{m}(m+1)}\exp\left(-\frac{\lambda\gamma KR^{2}}{m+1}\right)\right\} ,\label{eq:poutInfD}\end{equation}
 which gives a clear characterization for low-coherence routing selections.
As expected, multi-hop routing with the assistance of randomly deployed
relays improves the success probability by providing large potential
diversity, with the randomness in both relay locations and channel
states proving helpful.

We note that unlike the single hop scenario \cite{WebYanAnd2005},
our bound for outage probability without retransmissions is not globally
monotonically increasing with $\lambda$ if $D_{m}{\nrightarrow}\infty$.
For sufficiently large but not infinite $D_{m}$, the outage probability
can be approximated through a first-order Taylor expansion in the
low density regime: \begin{align}
 & p_{\text{out}}^{(m)}(\lambda)\nonumber \\
\geq & \exp\left\{ -\frac{G\kappa^{m}}{m+1}\left[\exp\left(-\frac{\Lambda R^{2}}{m+1}\right)-\exp\left(-{\Lambda D_{m}}\right)\right]\right\} \nonumber \\
\approx & ~1-{\lambda}\frac{G\kappa^{m}}{m+1}\left(D_{m}-\frac{R^{2}}{m+1}\right)\text{,}\end{align}
 which indicates large outage probability in the low density regime,
arising from the difficulty in guaranteeing a relay within range $d_{m}(Z_{m})\leq D_{m}$
in a sparse network. The detailed monotonicity can be more closely
examined by studying the function $f(\lambda)=\exp(-a\lambda)-\exp(-b\lambda)$
$(b>a>0)$, whose derivative can be computed as: \begin{align}
f'(\lambda)=\exp(-b\lambda)\left\{ b-a\exp\left[(b-a)\lambda\right]\right\} \text{.}\end{align}
 The maximum value of $f(\lambda)$ occurs at $\lambda_{0}=\frac{1}{b-a}\ln\frac{b}{a}$,
and $f(\lambda)$ is monotonically increasing at $(0,\lambda_{0}]$
and decreasing at $(\lambda_{0},\infty)$. Using this property, and
defining $\Delta=\frac{R^{2}}{(m+1)D_{m}}$, we can see that \begin{equation}
{\min}~p_{\text{out}}^{(m)}(\lambda)\geq\exp\left\{ \frac{G\kappa^{m}}{m+1}\left[\Delta{}^{\frac{1}{1-\Delta}}-\Delta{}^{\frac{\Delta}{1-\Delta}}\right]\right\} \text{,}\end{equation}
 where the minimizing $\lambda$ is: \begin{equation}
\lambda_{0}=\frac{1}{{\gamma}K\left(D_{m}-\frac{R^{2}}{m+1}\right)}\ln\frac{(m+1)D_{m}}{R^{2}}\text{.}\end{equation}
 Hence, $p_{\text{out}}^{(m)}(\lambda)$ is monotone in both $[0,\lambda_{0}]$
and $(\lambda_{0},\infty)$. Taking the inverse over $(\lambda_{0},\infty)$
will yield the bounds on maximum contention density.

\subsection{Transmission Capacity Upper Bound}

\label{sec:TCBound}

When $D_{m}\rightarrow\infty$, $\lambda_{0}$ goes to $0$. Therefore,
$p_{\text{out}}^{(m)}(\lambda)$ is monotonically increasing in $(0,\infty)$.
Therefore, we can get the following transmission capacity bound by
taking the inverse of this outage probability function.

\begin{corollary}\label{corollaryMultiRelay} (1) If each hop adopts
a single transmission, the transmission capacity can be bounded as
\begin{align}
T_{m}(\epsilon)~ & \leq~\frac{m\ln\frac{G{\pi}(1-\gamma)}{K{\gamma}}+{\ln}G-\ln(m+1)-\ln\ln\frac{1}{\epsilon}}{KR^{2}}(1-\epsilon)\nonumber \\
 & \overset{\Delta}{=}\text{ }T_{m}^{\text{ub}}(\epsilon)\text{,}\label{eqnTCMrelay}\end{align}
 where $\epsilon\geq\exp\left(-\frac{G\kappa^{m}}{m+1}\right)$.

(2) If best effort retransmissions is adopted with each hop utilizing
$k_{i}$ retransmissions, the transmission capacity can be bounded
as \begin{align}
T_{m}(\epsilon)~\leq~ & \left(\sum_{i=1}^{m+1}\frac{1}{k_{i}}\right)(1-\epsilon)\left\{ \frac{m\ln\frac{G{\pi}(1-\gamma)}{K{\gamma}}+{\ln}\left(Gm\right)}{KR^{2}\sum_{i=1}^{m+1}k_{i}}+\right.\nonumber \\
 & \quad\frac{-\ln\left[\left(\prod_{i=1}^{m+1}k_{i}\right)\left(\sum_{i=1}^{m+1}\frac{1}{k_{i}}\right)\right]-\ln\ln\frac{1}{\epsilon}}{KR^{2}\sum_{i=1}^{m+1}k_{i}}\text{,}\end{align}
 where $\epsilon\geq\exp\left(-\frac{mG\kappa^{m}}{\left(\prod_{i=1}^{m+1}k_{i}\right)\left(\sum_{i=1}^{m+1}\frac{1}{k_{i}}\right)}\right)$.
\end{corollary}

\begin{IEEEproof} See Appendix \ref{sec:Proof-of-CorollaryMultiRelay}.
\end{IEEEproof}

In the case where each hop use a single transmission, when $D_{m}\nrightarrow\infty$
but is reasonably large, the outage probability can be approximated
using L'Hôspital's rule \begin{align*}
\epsilon\geq\exp\left\{ -\frac{G\kappa^{m}}{m+1}\left[\exp\left(-\frac{\Lambda R^{2}}{m+1}\right)-\exp\left(-\frac{\Lambda D_{m}}{m+1}\right)\right]\right\} \text{.}\end{align*}
 By simple manipulation, the upper bound $T_{m}^{\text{ub}}(\epsilon,D_{m})$
on transmission capacity with $D_{m}$ constraint becomes \begin{align}
 & T_{m}^{\text{ub}}(\epsilon,D_{m})+\Theta\{\exp\{-T_{m}^{\text{ub}}(\epsilon,D_{m})K(m+1)D_{m}\}\}\nonumber \\
= & {T_{m}^{\text{ub}}}(\epsilon),\label{eqnTCwithD}\end{align}
 which means the gap between the general bound and the bound with
distance constraints will decay exponentially fast with $D_{m}$.

It should be noted that if we only impose a constraint on the maximum
number of allowable attempts $M$, it is difficult to get a closed-form
capacity bound. But since the outage bound is monotonically increasing
with $\lambda$, it would allow a numerical solution.

\section{Numerical Analysis and Discussion}

\label{sec:Discussion}

In this section we study the implications of the theoretical results
through simple numerical analysis and simulation. The presented plots
presume path-loss attenuation, Rayleigh fading, and no noise. The
SIR threshold $\beta$ is set to 1 while other parameters are varied.
The $\mathcal{S}-\mathcal{D}$ distance is primarily $R=4$ although
$R=6$ is also used. In a interference-limited environment the exact
values of $R$ and $\lambda$ are not particularly important since
the outage probability is constant for fixed $\lambda R^{2}$. In
the simulation, a spatial Poisson point process is generated. For
each spatial density $\lambda$, we pick $\mathcal{S}-\mathcal{D}$
pairs uniformly at random that has an average spatial density $\lambda\gamma$,
and let each $\mathcal{S}-\mathcal{D}$ pair perform uncoordinated
path selection. If there is any hop conflicting with routes selected
by other $\mathcal{S}-\mathcal{D}$ pair, then the transmission fails.
The set of connected paths from $\mathcal{S}$ to $\mathcal{D}$ is
determined through SIR measurements of each link, and constrained
to a maximum end-to-end distance of $D_{m}=60^{2}$ which qualitatively
approximates the extreme case of $D_{m}\rightarrow\infty$ for $R\leq6$.

\begin{figure}
\centering \includegraphics[width=3in]{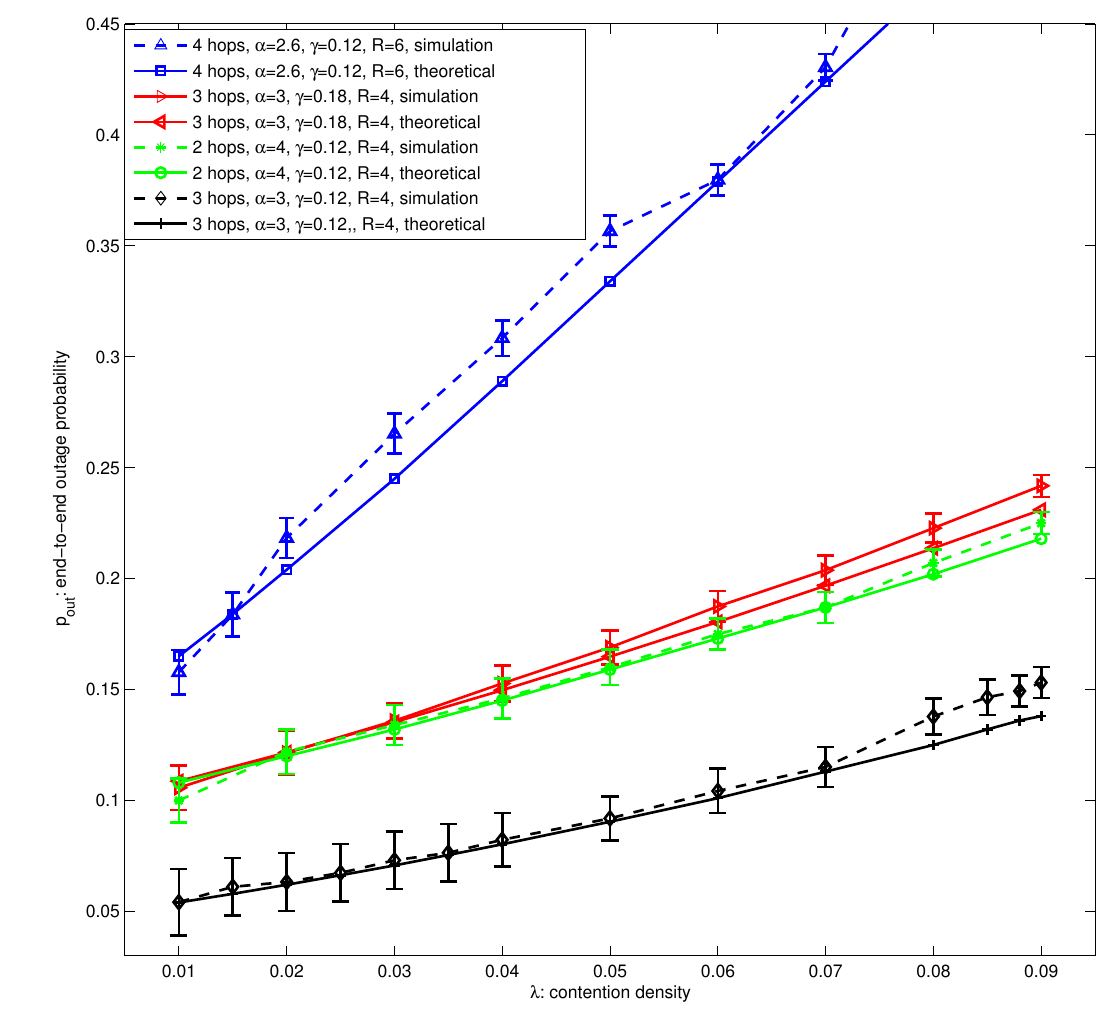}

\caption{The end-to-end outage probability $p_{\text{out}}$ computed numerically
and its lower bound versus the contention density $\lambda$ of all
potential transmitters. On each data point on the simulation curves,
an error bar indicating the confidence interval (which is chosen to
be the emperical standard deviation) is plotted. Note that the lower
bound is fairly tight for transmissions with $2$ hops and $3$ hops,
and becomes looser for transmissions through $4$ hops. }

\label{figSimPoutLambda234} %
\end{figure}

\begin{figure}
\centering \includegraphics[angle=-90,width=3.5in]{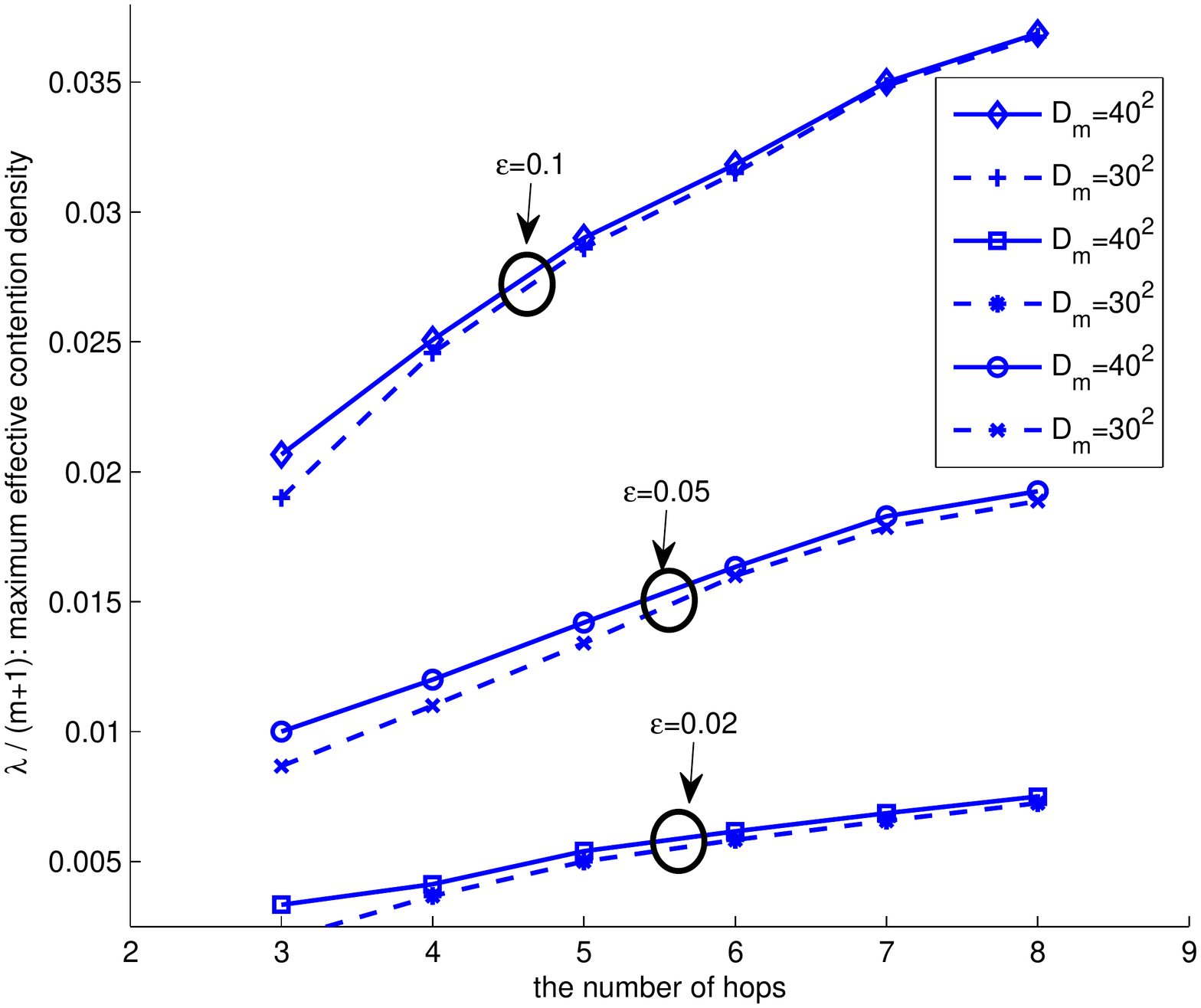}

\caption{The maximum allowable effective contention density $\lambda/(m+1)$
computed numerically versus the number of hops when $R=4$, $\alpha=3$,
$\beta=1$ for a typical $\mathcal{S}-\mathcal{D}$ pair. It can be
seen that the maximum effective density scales nearly linearly in
the number of hops. Note that increasing the distance constraint $D$
only provides fairly small throughput gain. }

\label{figSimCapacity} %
\end{figure}

\begin{figure}
\centering \includegraphics[width=2.5in,angle=-90]{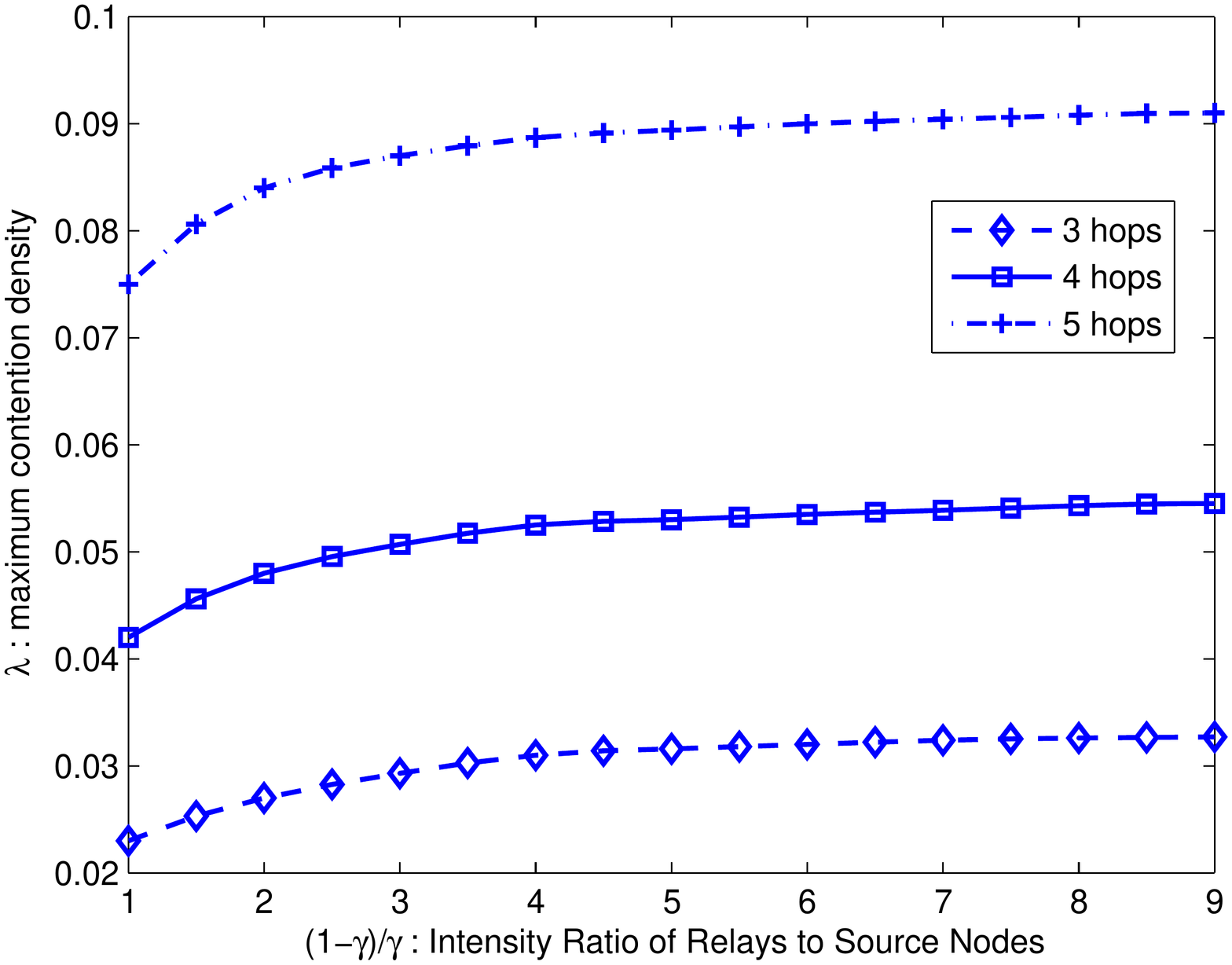}

\caption{The maximum allowable contention density $\lambda$ computed numerically
versus the intensity ratio of relays to source nodes $\frac{1-\gamma}{\gamma}$
when $\epsilon=0.05$ for a typical $\mathcal{S}-\mathcal{D}$ pair.
It can be observed that the maximum density scales logarithmically
in $\frac{1-\gamma}{\gamma}$.}

\label{figGammaCapacity} %
\end{figure}

\textbf{Tightness of outage lower bound.} The lower bound (\ref{eq:poutInfD})
is plotted against simulated outage probability in Fig. \ref{figSimPoutLambda234}.
The simulated outage probability takes into account the dependence
among consecutive transmissions and parallel path selection. For each
data point, an error bar is plotted to indicate the confidence interval
of the simulation results. Here, the width of the confidence interval
is chosen to be twice the empirical standard deviation. The bound
is observed to indeed be a lower bound and to be quite tight, albeit
slightly looser for increasing numbers of hops.

\textbf{The Number of Relays $m$.} Since $\Theta(\ln(m+1))$ is negligible
compared to $\Theta(m)$, the transmission capacity bound (\ref{eqnTCMrelay})
exhibits near linear scaling behavior with respect to the number of
relays $m$. This gain arises from the increasing route diversity
as $m$ grows, since more hops allows more potentially successful
routes. This gain does not depend on the noise level and is not achieved
by pre-determined routing approaches, which primarily are useful for
overcoming per-hop range limitations (i.e. noise). We caution that
this upper bound is likely to be increasingly optimistic for large
$m$, since longer potential routes will presumably result in higher
correlation between candidate paths. Fig. \ref{figSimCapacity} shows
the maximum allowable contention density versus the number of hops
for different outage constraints. As expected, the effective contention
density scales nearly linearly for small $m$ as expected, and then
diminishes rapidly for large $m$. In practice, a modest number of
hops would be taken since longer routes experience larger delay and
more protocol overhead. The proper choice of $m$ under realistic
correlation and protocol overhead models is an interesting topic for
future research. % This is different from the nearest neighbor routing analyzed in \cite{AndHaeKouWebJin2009} where increasing the number of hops is primarily used to change the noise-limited network into an interference-limited one. Instead, employing more hops provide throughput gain by exploiting path diversity through dynamic route selection rather than pre-determined approaches.

\textbf{Outage Probability Constraint ${\bf \epsilon}$.} The transmission
capacity bound is not sensitive to the outage constraint $\epsilon$
in the low outage regime, because the double logarithm as in $\ln\ln\frac{1}{\epsilon}$
largely reduces its sensitivity. For instance, when the target $\epsilon$
is decreased from $10^{-2}$ to $10^{-4}$, the throughput only experiences
a small constant loss. This is quite different than single-hop transmission
capacity, which exhibits linear scaling with $\epsilon$ in the low
outage regime and so going from $\epsilon=10^{-2}$ to $\epsilon=10^{-4}$
would in fact decrease the transmission capacity by two orders of
magnitude \cite{WebYanAnd2005}. Hence, multihop transmission capacity
is apparently much more robust to severe QoS constraints compared
to single hop.

\textbf{Availability of Relays.} Recall that nodes in the network
are divided into a fraction $\gamma$ that may transmit and $1-\gamma$
that are available as relays. Corollary \ref{corollaryMultiRelay}
implies that increasing the pool of relay nodes will logarithmically
increase throughput, so the diversity gains diminish rapidly once
a large enough pool to guarantee multi-hop route selection exists.
Note that we primarily consider fixed-portion relay models here, which
means the density of the pool of relays grows along with the density
of source nodes. Simulations in Fig. \ref{figGammaCapacity} show
the maximum contention density versus the intensity ratio $\frac{1-\gamma}{\gamma}$
of relays to source nodes, with $\epsilon=0.05$. The results can
be modified to study fixed-density relay models (where the density
of relay nodes is a fixed constant $\lambda_{r}$) by substituting
$\frac{1-\gamma}{\gamma}$ with $\lambda_{r}/\lambda_{t}$, which
we do not present here. 

\textbf{Sum-Squared-Distance Constraint} ${\bf D}_{m}$. The gap between
the distance-constrained maximum density $T_{m}(\epsilon,D_{m})$
and the transmission capacity $T_{m}(\epsilon)$ is subject to exponential
decay with respect to $(m+1)D_{m}$ as predicted in (\ref{eqnTCwithD}).
Hence, searching for multihop routes in a local region should be sufficient.
Fig. \ref{figSimCapacity} illustrates this when the $\mathcal{S}-\mathcal{D}$
distance is $R=4$. It can be observed that when $D$ is reasonably
large compared with $R^{2}$, increasing $D_{m}$ provides almost
no throughput gain. Also, this gain shrinks rapidly as $m$ increases,
which can also be expected from (\ref{eqnTCwithD}).

\textbf{Limitations and Future Directions.} The results of this paper
are well-suited to both fading and non-fading channels, but care should
be exercised in considering more diverse channel models like log-normal
shadowing, which do not necessarily lead to an exponential outage
probability expression \cite{WebAndJin2007}. In addition, the models
in this paper assume mutual independence among different links, which
would not hold in general, particularly for routing and scheduling
strategies that require cooperation. Furthermore, the theoretical
gap between our upper bound and the true capacity is unknown, and
how it grows with the number of hops is of interest.

The multihop transmission capacity bound shows that dynamic routing
selection is of significant importance when there is sufficient randomness
in the network as far as path gains and interference. In fact, predetermined
routing (like nearest-neighbor) is unlikely to approach the throughput
bound in interference-limited networks. A simple argument shows this.
Considering a typical source-destination pair, the outage probability
can be bounded as \begin{align}
1-p_{\text{out}}^{(m)} & =G\exp(-{\lambda}{\gamma}Kd_{m}(Z_{M}))\nonumber \\
 & \leq~G\exp\left(-\lambda\gamma K\frac{R^{2}}{m+1}\right).\end{align}
 The equality can be achieved if and only if the $m$ relays are equally
spaced along the line segment between source and destination. In fact,
from the properties of Poisson random process, this is almost surely
unlikely to occur, resulting in a strict inequality. Setting ${\lambda}{\gamma}(1-\epsilon)/(m+1)$
to $T_{m}^{\text{ub}}(\epsilon)$, we can immediately get an upper
bound \begin{align}
T_{m}^{\text{ub}}(\epsilon)=\frac{1-\epsilon}{KR^{2}}\ln{\frac{G}{1-\epsilon}},\end{align}
 which is exactly equal to the single hop case. This suggests that
predetermined routing will not provide further throughput gain in
interference-limited networks compared with single hop direct transmission. 

We note that the power-limited regime (i.e. including noise) is not
considered in this paper. Although noise is unimportant in the high
density regime, it can be quite important in the low density regime,
which is often power-limited. Our framework is primarily based upon
an exponential form of per-hop success probability, which does not
hold in the low SNR case. From a capacity perspective, the high density
case is of more interest, since in the low density (power-limited)
regime, nodes have far fewer options as far as selecting relays, and
spatial reuse in the network is not very important. In fact, multihopping
is known to be particularly helpful in changing a power-limited network
to an interference-limited one by increasing the SNR in each hop,
consistent with\cite{AndHaeKouWebJin2009,SikLanHaeCosFuj06}. 

The design of transmission strategies that exploits the path diversity
gain are left for future work. We conjecture that hop-by-hop route
selection --- which is much more realistic in a distributed network
than the complete route selection assumed here --- will achieve a
lower diversity order (and hence transmission capacity).

\appendices{}

\section{Single hop success probability under Nakagami fading }

\label{AppendixNaka} The single-hop success probability with Nakagami
fading can be developed as \begin{align}
g_{0}(r_{ij},\lambda_{t}) & =\int_{0}^{\infty}\mathbb{P}\left(\frac{zr_{ij}^{-\alpha}}{t}\geq\beta\right)f_{I_{\Phi}}(t)dt\nonumber \\
 & =\sum_{k=0}^{m_{0}-1}\frac{(-m_{0}\beta r_{ij}^{\alpha})^{k}}{k!}\mathcal{L}_{I_{\Phi}}^{(k)}(m_{0}{\beta}r_{ij}^{\alpha})\text{,}\end{align}
 where $\mathcal{L}_{I_{\Phi}}(s)$ is the Laplace transform of the
general Poisson shot noise process, and $\mathcal{L}_{I_{\Phi}}^{(k)}(s)$
denotes the $k^{\text{th}}$ derivative of $\mathcal{L}_{I_{\Phi}}(s)$.
The closed-form formulas of them are given by \cite{HunAndWeb2008}
as \begin{align}
\mathcal{L}_{I_{\Phi}}(s)=\exp\left\{ -\lambda_{t}\Omega_{m_{0}}\left(\frac{s}{m_{0}}\right)^{\frac{2}{\alpha}}\right\} ,\end{align}
 \begin{align}
 & \mathcal{L}_{I_{\Phi}}^{(k)}(s)\nonumber \\
= & \frac{\exp\left\{ -\lambda_{t}\Omega_{m_{0}}\left(\frac{s}{m_{0}}\right)^{\frac{2}{\alpha}}\right\} }{(-s)^{k}}\sum_{j=1}^{k}\left[\frac{-2\lambda_{t}\Omega_{m_{0}}}{\alpha}\left(\frac{s}{m_{0}}\right)^{\frac{2}{\alpha}}\right]^{j}\Upsilon_{k,j},\end{align}
 where $\Upsilon_{k,j}$ is a constant defined in \cite{HunAndWeb2008},
and \begin{align}
\Omega_{m_{0}}=\frac{2\pi}{\alpha}\sum_{k=0}^{m_{0}-1}\binom{m}{k}B(k+\frac{2}{\alpha},m_{0}-k-\frac{2}{\alpha})\end{align}
 with $B(a,b)$ denoting Beta function. By manipulation, we have \begin{align*}
g_{0}(r_{ij},\lambda_{t}) & =\exp\left\{ -\lambda_{t}\Omega_{m_{0}}{\beta}^{\frac{2}{\alpha}}r_{ij}^{2}\right\} \cdot\\
 & \left\{ 1+\sum_{k=1}^{m_{0}-1}\sum_{l=1}^{k}\frac{1}{k!}\left[-\frac{2\lambda_{t}\Omega_{m_{0}}{\beta}^{\frac{2}{\alpha}}r_{ij}^{2}}{\alpha}\right]^{l}\Upsilon_{k,l}\right\} .\end{align*}

Generally speaking, this does not have an expected exponential form.
But we can simplify the expression in certain cases. For small single-hop
outage constraint $\epsilon$, we have $\lambda_{t}\Omega_{m_{0}}{\beta}^{\frac{2}{\alpha}}r_{ij}^{2}{\ll}1$,
therefore $g_{0}(r_{ij},\lambda_{t})$ can be approximated as \begin{align}
g_{0}(r_{ij},\lambda_{t})\approx\exp\left\{ -\lambda_{t}\Omega_{m_{0}}\beta^{\frac{2}{\alpha}}r_{ij}^{2}\right\} .\end{align}
 In contrast, for large single-hop outage regime, i.e. $\lambda_{t}\Omega_{m_{0}}\beta^{\frac{2}{\alpha}}r_{ij}^{2}\gg1$,
employing L'Hospital's rule yields \begin{align*}
 & g_{0}(r_{ij},\lambda_{t})\\
\approx & \left\{ 1+\sum_{k=1}^{m_{0}-1}\sum_{l=1}^{k}\frac{l!}{k!}\left(-\frac{2}{\alpha}\right)^{l}\Upsilon_{k,l}\right\} \exp\left\{ -\lambda_{t}\Omega_{m_{0}}\beta^{\frac{2}{\alpha}}r_{ij}^{2}\right\} .\end{align*}
 We summarize them as follows \begin{align*}
\text{low outage regime: } & K_{\text{NF}}=\Omega_{m_{0}}\beta^{\frac{2}{\alpha}},~G_{\text{NF}}=1,\\
\text{high outage regime: } & K_{\text{NF}}=\Omega_{m_{0}}\beta^{\frac{2}{\alpha}},\\
 & G=1+\sum_{k=1}^{m_{0}-1}\sum_{l=1}^{k}\frac{l!}{k!}\left(\frac{-2}{\alpha}\right)^{l}\Upsilon_{k,l}\end{align*}
 Since practical system typically require low outage probability,
our analysis may still work to a certain extent.

\section{\label{sec:Proof-of-Thm-1}Proof of Theorem \ref{thmGeneralOutBound}}

Let the high-dimensional feasible region $\mathcal{F}$ for relay
sets be the allowable range to select relays from determined by different
routing protocols and design parameters. Denote by $\mathcal{A}$
the event that there is \textit{{no}} relay set within $\mathcal{F}$
that can successfully complete forwarding. Ignoring the edge effect,
we attempt to approximately divide $\mathcal{F}$ into $n$ disjoint
hypercubes $\mathcal{F}_{i}(1{\leq}i{\leq}n)$ each of equal volume.
For sufficiently large $n$, this approximation is exact. Let $\mathcal{A}_{i}(1{\leq}i{\leq}n)$
be the event that there exists no potential relay set within $\mathcal{F}_{i}$
that can complete forwarding. Since the outage event $\mathcal{A}$
occurs only when there is no potential relay set in any of the region
$\mathcal{F}_{i}$, we have $\mathcal{A}=\bigcap_{i=1}^{n}\mathcal{A}_{i}$.
Consider the hypercube $\mathcal{F}_{i}$ as $[(x_{1},y_{1},\dots,x_{m},y_{m}),(x_{1}+{\delta}x_{1},y_{1}+{\delta}y_{1},\dots,x_{m}+{\delta}x_{m},y_{m}+{\delta}y_{m})]$
when ${\delta}x_{i}{\rightarrow}0$ and ${\delta}y_{i}{\rightarrow}0$.
Define $Z_{i}=(x_{1},y_{1},\dots,x_{m},y_{m})$. Since this is a simple
point process, we can approximate the void probability as follows
if the Lebesgue measure $v_{2m}(\mathcal{F}_{i})$ is small or $n$
is sufficiently large \begin{align}
 & \lim_{{\delta}x_{i}{\rightarrow}0,{\delta}y_{i}{\rightarrow}0}\mathbb{P}(\mathcal{A}_{i})\nonumber \\
= & ~\lim_{{\delta}x_{i}{\rightarrow}0,{\delta}y_{i}{\rightarrow}0}1-g_{m}(Z_{i},{\lambda_{t}})\prod_{i=1}^{m}\left(1-\exp(-\tilde{\lambda}{\delta}x_{i}{\delta}y_{i})\right)\\
= & ~\lim_{{\delta}x_{i}{\rightarrow}0,{\delta}y_{i}{\rightarrow}0}1-g_{m}(Z_{i},{\lambda_{t}})\prod_{i=1}^{m}\tilde{\lambda}{\delta}x_{i}{\delta}y_{i}\nonumber \\
= & ~\lim_{{\delta}x_{i}{\rightarrow}0,{\delta}y_{i}{\rightarrow}0}\exp\left(-\tilde{\lambda}^{m}g_{m}(Z_{i},{\lambda_{t}})v_{2m}(\mathcal{F}_{i})\right).\end{align}
 Consider two realizations $\omega$ and $\omega'$ of this higher-dimensional
point process, and denote $\omega\preceq\omega'$ if $\omega'$ can
be obtained from $\omega$ by adding points. An event $\mathcal{A}_{i}$
is said to be increasing if for every $\omega\preceq\omega'$, $\mathbb{I}_{\mathcal{A}_{i}}\left(\omega\right)\leq\mathbb{I}_{\mathcal{A}_{i}}\left(\omega'\right)$
with $\mathbb{I}_{\mathcal{A}_{i}}$ denoting the indicator function
of the event $\mathcal{A}_{i}$. If $\mathcal{A}_{i}(1{\leq}i{\leq}n)$
are all increasing events, then the Harris-FKG inequality \cite{MeeRoy}
yields \begin{equation}
\mathbb{P}(\mathcal{A})=\mathbb{P}(\bigcap_{i=1}^{n}\mathcal{A}_{i})\geq\prod_{i=1}^{n}\mathbb{P}(\mathcal{A}_{i}).\end{equation}
 Letting $n$ go to infinity, we can get the lower bound of outage
probability as follows \begin{align}
\mathbb{P}(\mathcal{A}) & \geq\lim_{n\rightarrow\infty}\prod_{i=1}^{n}\mathbb{P}(\mathcal{A}_{i})\nonumber \\
 & =\exp\left(-\tilde{\lambda}^{m}g_{m}(Z_{i},{\lambda_{t}})\lim_{n\rightarrow\infty}\sum_{i=1}^{n}v_{2m}(\mathcal{F}_{i})\right)\nonumber \\
 & =\exp(-\mathbb{E}(N_{m})).\end{align}

\section{\label{sec:Proof-of-Lemma-AvgMultiRelay}Proof of Lemma \ref{thmAvgMultiRelay}}

The isosurface of $d_{m}(Z_{m})=a$ has the following coordinate geometry
form \begin{align}
X_{\text{sum}}+Y_{\text{sum}}~=~a\text{,}\label{eqnQuadratic}\end{align}
 where \begin{align*}
X_{\text{sum}}~ & =~\left(x_{1}+\frac{R}{2}\right)^{2}+\sum_{i=1}^{m-1}(x_{i+1}-x_{i})^{2}+\left(x_{m}-\frac{R}{2}\right)^{2}\text{,}\\
Y_{\text{sum}}~ & =~y_{1}^{2}+\sum_{i=1}^{m-1}(y_{i+1}-y_{i})^{2}+y_{m}^{2}\text{.}\end{align*}

If we treat $x_{i},y_{i}~(1\leq i\leq m)$ as mutually orthogonal
coordinates, then (\ref{eqnQuadratic}) forms a quadratic surface
in $2m$-dimensional space. See Fig. \ref{figSimPoutLambda234} for
an illustration when $m=1$. From the properties of quadratic forms,
the $x$ part and $y$ part of (\ref{eqnQuadratic}) can be expressed
as \begin{align}
\left\{ \begin{aligned}X_{\text{sum}}~ & =~{\bf (CX-{\bf R_{x})^{T}\Lambda_{x}(CX-R_{x})}}+t_{m}R^{2}\text{,}\\
Y_{\text{sum}}~ & =~{\bf (\tilde{C}Y)^{T}\Lambda_{y}(\tilde{C}Y)},\end{aligned}
\right.\end{align}
 where ${\bf {C},\tilde{C}}$ are orthogonal matrices, ${\bf \Lambda_{x},\Lambda_{y}}$
are diagonal matrices, ${\bf {R_{x}}}$ is a $m$-dimensional vector,
and $t_{m}$ is a constant that will be determined in the sequel.
Here, the orthogonal transformation of ${\bf X~(Y)}$ by ${\bf C~(\tilde{C})}$
and translation transformation by ${\bf R_{x}}$ only result in rotation,
flipping or translation of the quadratic surface without changing
the shape of it. Since the corresponding quadratic terms of $X_{\text{sum}}$
and $Y_{\text{sum}}$ have equivalent coefficients, we have ${\bf \Lambda}_{m}\overset{\Delta}{=}{\bf \Lambda_{x}}={\bf \Lambda_{y}}$.
Denote the symmetric quadratic-form matrix corresponding to $Y_{\text{sum}}$
as ${\bf {A}}_{m}$, then ${\bf {A}}_{m}$ is the following tridiagonal
matrix of dimension $m$: \begin{align}
{\bf {A}}_{m}=\begin{pmatrix}2 & -1 & 0 & \dots & 0\\
-1 & 2 & -1 & \dots & 0\\
0 & -1 & 2 & \dots & 0\\
\vdots & \vdots & \vdots &  & \vdots\\
0 & 0 & 0 & \dots & 2\end{pmatrix}\text{.}\end{align}
 In fact, ${\bf \Lambda}_{m}$ is the canonical form of ${\bf {A}}_{m}$
with its eigenvalues on the main diagonal. Through orthogonal transformation
and translation, $X_{\text{sum}},Y_{\text{sum}}$ can be brought to
the explicit form: \begin{equation}
\left\{ \begin{aligned}X_{\text{sum}} & =\sum_{i=1}^{m}\lambda_{i}\tilde{x_{i}}^{2}~+~t_{m}R^{2}\text{,}\\
Y_{\text{sum}} & =\sum_{i=1}^{m}\lambda_{i}\tilde{y_{i}}^{2},\end{aligned}
\right.\end{equation}
 where $\tilde{x_{i}},\tilde{y_{i}}$ are the new orthogonal coordinates
and $\lambda_{i}$ is the $i$th eigenvalue of ${\bf {A}}_{m}$. By
its definition, $X_{\text{sum}}$ is positive definite, and the following
minimum value can be obtained if and only if $m$ relays are placed
equidistant along the line segment between the source and destination:
\begin{align}
X_{\text{sum}}~\geq & ~\frac{(|\frac{R}{2}+x_{1}|+|x_{2}-x_{1}|+\dots+|x_{m}-\frac{R}{2}|)^{2}}{m+1}\nonumber \\
\geq & ~\frac{R^{2}}{m+1}\text{.}\end{align}
 Therefore, $t_{m}=\frac{1}{m+1}$. Now, (\ref{eqnQuadratic}) can
be brought to: \begin{equation}
\sum_{i=1}^{m}\lambda_{i}\tilde{x_{i}}^{2}~+~\sum_{i=1}^{m}\lambda_{i}\tilde{y_{i}}^{2}~=~a-\frac{R^{2}}{m+1}\text{.}\end{equation}

\begin{figure}
\centering \includegraphics[width=3in]{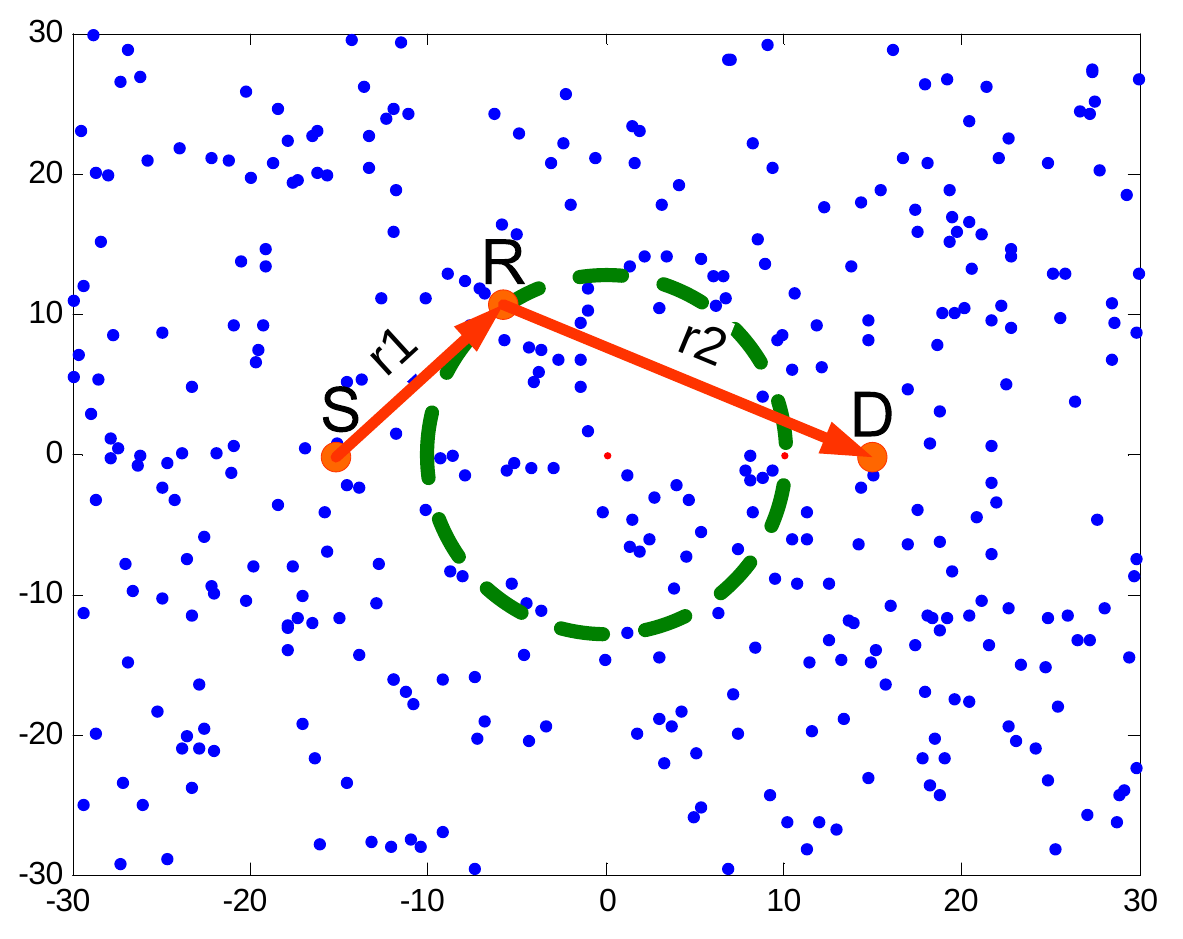}

\caption{In single relay scenario, the contour of end-to-end success probability
$g_{1}=G^{2}\exp\left(-\lambda_{t}K\left(r_{1}^{2}+r_{2}^{2}\right)\right)$
is exactly a circle. The plot is a realization of Poisson point process
of $\lambda=0.1/\text{unit area}$. The source $S$ and destination
$D$ are a distance $30$ apart with relay $R$ on the dotted circle
satisfying $r_{1}^{2}+r_{2}^{2}=500$. }

\label{figEllipsoid-1} %
\end{figure}

From the positive definiteness of ${\bf {A}}_{m}$, ${\bf {\lambda_{i}}}~>0$
holds for all $i$, i.e. the above equation forms the surface of a
$2m$-dimensional ellipsoid. Fig. \ref{figEllipsoid-1} illustrates
the ellipsoid when $m=1$, which reduces to a circle. The Lebesgue
measure of the ellipsoid can be written as: \begin{equation}
V_{m}(a)=\frac{\pi^{m}(a-\frac{R^{2}}{m+1})^{m}}{m!\prod_{i=1}^{m}\lambda_{i}}=\frac{\pi^{m}(a-\frac{R^{2}}{m+1})^{m}}{m!\det\left({\bf A}_{m}\right)}\text{.}\end{equation}
 We also need to determine $\det({\bf A}_{m})$, which can be computed
by the Laplace expansion of the determinant \begin{equation}
\det({\bf A}_{m})=2\det({\bf A}_{m-1})-\det({\bf A}_{m-2})\text{.}\end{equation}
 Solving this recursive form with the initial value $\det({\bf A}_{1})=2$
and $\det({\bf A}_{2})=3$ yields \begin{equation}
\det({\bf A}_{m})=m+1~~\Rightarrow~~V_{m}(a)=\frac{\pi^{m}(a-\frac{R^{2}}{m+1})^{m}}{(m+1)!}\text{.}\end{equation}

Now, we can compute the outage probability. Integrating over different
isosurfaces with $g(Z_{m},\lambda\gamma)=\exp(-\lambda\gamma Ka)$,
and defining $H=\frac{\pi(1-\gamma)}{\gamma K}$, we can compute the
average number of potential relay sets as \begin{align}
 & \mathbb{E}(N_{m})~\nonumber \\
= & ~\int_{\frac{R^{2}}{m+1}}^{D_{m}}\lambda^{m}(1-\gamma)^{m}\frac{\mathrm{d}V_{m}(a)}{\mathrm{d}a}G^{m+1}\exp(-\Lambda a)\mathrm{d}a\\
= & ~G^{m+1}\lambda^{m}(1-\gamma)^{m}\cdot\nonumber \\
 & \quad\int_{\frac{R^{2}}{m+1}}^{D_{m}}\frac{m\pi^{m}(a-\frac{R^{2}}{m+1})^{m-1}}{(m+1)!}\exp(-\Lambda a)\mathrm{d}a\\
= & ~\frac{mG^{m+1}H^{m}\exp(-\frac{\Lambda R^{2}}{m+1})}{(m+1)!}\int_{0}^{\Lambda(D_{m}-\frac{R^{2}}{m+1})}x^{m-1}e^{-x}\mathrm{d}x\label{eqnGammaFunc}\\
= & ~\frac{mG^{m+1}H^{m}e^{-\frac{\Lambda R^{2}}{m+1}}}{(m+1)!}\left\{ e^{-x}\sum\limits _{0}^{m-1}\frac{(m-1)!}{i!}x^{i}\right\} _{\Lambda(D_{m}-\frac{R^{2}}{m+1})}^{0}\nonumber \\
= & ~\frac{G^{m+1}H^{m}}{m+1}\left\{ \exp(-\frac{\Lambda R^{2}}{m+1})-\right.\nonumber \\
 & \quad\quad\exp(-\Lambda D_{m})\sum\limits _{0}^{m-1}\frac{1}{i!}\left(\Lambda(D_{m}-\frac{R^{2}}{m+1})\right)^{i}\text{.}\nonumber \end{align}

It is worth noting that a relay set may contain the same location
for different relays. This can be interpreted as employing the same
node in different frequency bands for forwarding, although this is
not common in practical routing. We notice that these sets form $\binom{m}{2}$
hyperplanes in the $2m$ dimensional hyperspace, which are of measure
$0$. Hence, even if we require distinct relays and take the integral
over feasible regions, we will still get the same results.

\section{\label{sec:Proof-of-Lemma}Proof of Lemma \ref{lemmaRetran}}

We proceed in a similar spirit as in the proof of Lemma \ref{thmAvgMultiRelay}.
Define ${\bf n}=(n_{1},\cdots,n_{m+1})^{T}$ and ${\bf k}=\left(k_{1},\cdots,k_{m+1}\right)^{T}$.
When the $i$th hop is of distance $r_{i}$ and $k_{i}$ attempts
are employed in the $i$th hop, the probability for successful reception
is given by\begin{align}
 & g\left(r_{1},k_{1},\cdots,r_{m+1},k_{m+1}\right)\nonumber \\
= & \prod_{i=1}^{m+1}\left[1-\left(1-G\exp\left(-\lambda_{t}Kr_{i}^{2}\right)\right)^{k_{i}}\right]\nonumber \\
= & \prod_{i=1}^{m+1}\left[\sum_{n_{i}=1}^{k_{i}}\left(-1\right)^{n_{i}+1}\left(\begin{array}{c}
k_{i}\\
n_{i}\end{array}\right)G^{n_{i}}\exp\left(-\lambda_{t}Kn_{i}r_{i}^{2}\right)\right]\nonumber \\
= & \sum_{{\bf n}:\text{ }{\bf 1}\preceq{\bf n}\preceq{\bf k}}\left(-1\right)^{m+1+\sum_{i=1}^{m+1}n_{i}}G^{\sum_{i=1}^{m+1}n_{i}}\nonumber \\
 & \quad\quad\quad\exp\left(-\lambda_{t}K\sum_{i=1}^{m+1}n_{i}r_{i}^{2}\right)\prod_{i=1}^{m+1}\left(\begin{array}{c}
k_{i}\\
n_{i}\end{array}\right),\label{eq:SuccessProbabilityMultiAttempt}\end{align}
 where ${\bf 1}:=\left(1,\cdots,1\right)^{T}$. Therefore, we redefine
$X_{\text{sum}}$, $Y_{\text{sum}}$ to be \begin{align*}
X_{\text{sum}}^{*}~ & =~n_{1}\left(x_{1}+\frac{R}{2}\right)^{2}+\sum_{i=1}^{m-1}n_{i+1}(x_{i+1}-x_{i})^{2}\\
 & \quad\quad+n_{m+1}\left(x_{m}-\frac{R}{2}\right)^{2}\text{,}\\
Y_{\text{sum}}^{*}~ & =~n_{1}y_{1}^{2}+\sum_{i=1}^{m-1}n_{i+1}(y_{i+1}-y_{i})^{2}+n_{m+1}y_{m}^{2}\text{.}\end{align*}
 The Cauchy-Schwartz inequality indicates \begin{align*}
\left(\sum_{i=1}^{m+1}n_{i}r_{i}^{2}\right)\left(\sum_{i=1}^{m+1}\frac{1}{n_{i}}\right)\geq\left(\sum_{i=1}^{m+1}r_{i}\right)^{2}=R^{2},\end{align*}
 \begin{align}
\Rightarrow~~X_{\text{sum}}^{*}\geq\frac{R^{2}}{\sum_{i=1}^{m+1}\frac{1}{n_{i}}}.\end{align}
 Therefore, the Lebesgue measure of the ellipsoid $d_{m}^{*}(Z_{m}){\leq}a$
can be calculated as \begin{align}
V_{m}(a)^{*}=\frac{\pi^{m}\left(a-\frac{R^{2}}{\sum_{i=1}^{m+1}1/n_{i}}\right)^{m}}{m!\det\left({\bf A}_{m}^{*}\right)},\end{align}
 where ${\bf A}_{m}^{*}$ is the canonical form corresponding to $Y_{\text{sum}}^{*}$
and can be written as \begin{align*}
{\bf A}_{m}^{*}=\begin{pmatrix}n_{1}+n_{2} & -n_{2} & 0 & \dots & 0\\
-n_{2} & n_{2}+n_{3} & -n_{3} & \dots & 0\\
0 & -n_{3} & n_{3}+n_{4} & \dots & 0\\
\vdots & \vdots & \vdots &  & \vdots\\
0 & 0 & 0 & \dots & n_{m}+n_{m+1}\end{pmatrix}\text{.}\end{align*}
 By Laplace expansion of the determinant, we get \begin{align*}
\det({\bf A}_{m}^{*})=(n_{m}+n_{m+1})\det({\bf A}_{m-1}^{*})-n_{m}^{2}\det({\bf A}_{m-2}^{*}),\end{align*}
 \begin{align}
\Rightarrow~\det({\bf A}_{m}^{*})=\left(\prod_{i=1}^{m+1}n_{i}\right)\left(\sum_{i=1}^{m+1}\frac{1}{n_{i}}\right),\end{align}
 which follows by induction. Define $a=\sum_{i=1}^{m+1}n_{i}r_{i}^{2}$.
Taking an integral over different isosurfaces yields\begin{align}
h_{{\bf n}}\overset{\Delta}{=} & ~\int_{\frac{R^{2}}{\sum_{i=1}^{m+1}1/n_{i}}}^{\infty}\frac{\mathrm{d}V_{m}^{*}(a)}{\mathrm{d}a}\exp\left(-\Lambda a\right)\mathrm{d}a\nonumber \\
= & ~\int_{\frac{R^{2}}{\sum_{i=1}^{m+1}1/n_{i}}}^{\infty}\frac{\pi^{m}\left(a-\frac{R^{2}}{\sum_{i=1}^{m+1}1/n_{i}}\right)^{m-1}}{(m-1)!\left(\prod_{i=1}^{m+1}n_{i}\right)\left(\sum_{i=1}^{m+1}\frac{1}{n_{i}}\right)}e^{-\Lambda a}\mathrm{d}a\nonumber \\
= & ~\frac{\pi^{m}e^{-\frac{\Lambda R^{2}}{\sum_{i=1}^{m+1}1/k_{i}}}\int_{0}^{\infty}x^{m-1}e^{-x}\mathrm{d}x}{\Lambda^{m}(m-1)!\left(\prod_{i=1}^{m+1}n_{i}\right)\left(\sum_{i=1}^{m+1}\frac{1}{n_{i}}\right)}\nonumber \\
= & ~\frac{\pi^{m}m\exp\left(-\frac{\Lambda R^{2}}{\sum_{i=1}^{m+1}1/k_{i}}\right)}{\Lambda^{m}\left(\prod_{i=1}^{m+1}n_{i}\right)\left(\sum_{i=1}^{m+1}\frac{1}{n_{i}}\right)}.\label{eq:IntegralMultiAttempt}\end{align}

By combining (\ref{eq:SuccessProbabilityMultiAttempt}) and (\ref{eq:IntegralMultiAttempt}),
we can derive the average number of relay sets when retransmitting
$k_{i}$ times in the $i^{\text{th}}$ hop as \begin{align}
 & \mathbb{E}_{{\bf {k}}}(N_{m})\nonumber \\
= & \lambda^{m}(1-\gamma)^{m}\text{ }\sum_{{\bf n}:\text{ }{\bf 1}\preceq{\bf n}\preceq{\bf k}}\left(-1\right)^{m+1}\cdot\nonumber \\
 & \quad\quad\left(-G\right)^{\sum_{j=1}^{m+1}n_{j}}\prod_{i=1}^{m+1}\left(\begin{array}{c}
k_{i}\\
n_{i}\end{array}\right)h_{{\bf n}}\nonumber \\
= & \lambda^{m}(1-\gamma)^{m}\sum_{{\bf n}:\text{ }{\bf 1}\preceq{\bf n}\preceq{\bf k}}\left(-1\right)^{m+1}\left(-G\right)^{\sum_{j=1}^{m+1}n_{j}}\cdot\nonumber \\
 & \quad\quad\prod_{i=1}^{m+1}\left(\begin{array}{c}
k_{i}\\
n_{i}\end{array}\right)\frac{\pi^{m}m\exp\left(-\frac{\Lambda R^{2}}{\sum_{j=1}^{m+1}1/k_{j}}\right)}{\Lambda^{m}\left(\prod_{i=1}^{m+1}n_{i}\right)\left(\sum_{i=1}^{m+1}\frac{1}{n_{i}}\right)}\nonumber \\
= & \frac{\left(-1\right)^{m+1}\pi^{m}(1-\gamma)^{m}}{\gamma^{m}K^{m}}\sum_{{\bf n}:\text{ }{\bf 1}\preceq{\bf n}\preceq{\bf k}}\left(-G\right)^{\sum_{j=1}^{m+1}n_{j}}\cdot\nonumber \\
 & \quad\quad\prod_{i=1}^{m+1}\left(\begin{array}{c}
k_{i}\\
n_{i}\end{array}\right)\frac{m\exp\left(-\frac{\Lambda R^{2}}{\sum_{j=1}^{m+1}1/k_{j}}\right)}{\left(\prod_{i=1}^{m+1}n_{i}\right)\left(\sum_{i=1}^{m+1}\frac{1}{n_{i}}\right)}.\label{eq:EkNm}\end{align}
 In addition, we can impose a constraint on the maximum total number
of attempts $M$ without specifying the number of transmissions for
each hop. For a typical relay set with the $i^{\text{th}}$ hop of
distance $r_{i}$, we denote by $p_{i}$ the success probability of
hop $i$ in any time slot. Among these $M$ time slots, successful
reception occurs when there exists $m+1$ slots $t_{i}(1\leq i\leq m+1)$
that satisfy: (1) transmission in the $i^{\text{th}}$ hop is successful
at time $t_{i}$; (2) for $1\leq i<j\leq m+1$, we have $1\leq t_{i}<t_{j}\leq M$.
We apply a greedy approach to search for all possible scenarios that
allow successful reception, which can be determined by the smallest
$t=(t_{1},\cdots,t_{m+1})$ that satisfies the above two requirements.
By {}``smallest'' we mean there is no $\tilde{t}\preceq t$ that
meets the requirement. This is identical to finding the interval $(t_{1}-1,t_{2}-t_{1}-1,\cdots,t_{m+1}-t_{m}-1)$,
or equivalently, finding a vector ${\bf {j}}=(j_{1},\cdots,j_{m+1})$
such that ${\bf {j}}\succeq0$ and ${\bf {j}}^{T}\cdot{\bf {1}}\leq M-m-1$.
Hence, the success probability with $m+1$ hop routing can be calculated
as \begin{align}
 & g_{(m,M)}(r_{1},\cdots,r_{m+1})\nonumber \\
= & \prod_{i=1}^{m+1}p_{i}\left\{ \sum_{\substack{{\bf {j}}^{T}\cdot{\bf {1}}\leq M-m-1\\
{\bf {j}}\succeq0}
}\prod_{i=1}^{m+1}(1-p_{i})^{j_{i}}\right\} \\
= & \left(\prod_{i=1}^{m+1}p_{i}\right)\sum_{\substack{{\bf {j}}^{T}\cdot{\bf {1}}\leq M-m-1\\
{\bf {j}}\succeq0}
}\sum_{0\preceq{\bf {k}}\preceq{\bf {j}}}(-1)^{{\bf {k}}^{T}\cdot{\bf {1}}}\prod_{l=1}^{m+1}{j_{l} \choose k_{l}}\prod_{i=1}^{m+1}p_{i}^{k_{i}}\nonumber \\
= & \sum_{\substack{{\bf {k}}^{T}\cdot{\bf {1}}\leq M-m-1\\
{\bf {k}}\succeq0}
}(-1)^{{\bf {k}}^{T}\cdot{\bf {1}}}\sum_{\substack{{\bf {j}}\succeq{\bf {k}}\\
{\bf {j}}^{T}\cdot{\bf {1}}\leq M-m-1}
}\prod_{l=1}^{m+1}{j_{l} \choose k_{l}}\prod_{i=1}^{m+1}p_{i}^{k_{i}+1}\nonumber \\
= & \sum_{\substack{{\bf {k}}^{T}\cdot{\bf {1}}\leq M\\
{{\bf {k}}\succeq{\bf {1}}}}
}(-1)^{{\bf {k}}^{T}\cdot{\bf {1}}}\sum_{\substack{{\bf {j}}^{T}\cdot{\bf {1}}\leq M\\
{\bf {j}}\succeq{\bf {k}}}
}\prod_{l=1}^{m+1}{j_{l}-1 \choose k_{l}-1}\prod_{i=1}^{m+1}p_{i}^{k_{i}}.\end{align}
 Therefore, we can obtain \begin{align*}
\mathbb{E}_{M}(N_{m}) & =\sum_{\substack{{\bf {k}}^{T}\cdot{\bf {1}}\leq M\\
{{\bf {k}}\succeq{\bf {1}}}}
}(-1)^{{\bf {k}}^{T}\cdot{\bf {1}}}\sum_{\substack{{\bf {j}}^{T}\cdot{\bf {1}}\leq M\\
{\bf {j}}\succeq{\bf {k}}}
}\prod_{l=1}^{m+1}{j_{l}-1 \choose k_{l}-1}E_{{\bf {k}}}(N_{m}),\end{align*}
 where $\mathbb{E}_{{\bf {k}}}(N_{m})$ is given in (\ref{eq:EkNm}).

\section{\label{sec:Proof-of-CorollaryMultiRelay}Proof of Corollary \ref{corollaryMultiRelay}}

Let $D_{m}\rightarrow\infty$, then the integral part in (\ref{eqnGammaFunc})
becomes a Gamma function: \begin{align*}
\lim_{D_{m}\rightarrow\infty}\int_{0}^{{\lambda\gamma}K(D_{m}-\frac{R^{2}}{m+1})}x^{m-1}e^{-x}\mathrm{d}x=\Gamma(m)=(m-1)!\end{align*}
 By setting the outage probability to be $\epsilon$, we can simplify
(\ref{eqnPoutMRelay}) as \begin{equation}
\epsilon\geq\exp\left\{ -\frac{G^{m+1}\pi^{m}(1-\gamma)^{m}}{\gamma^{m}K^{m}(m+1)}\exp(-\frac{\lambda\gamma}{m+1}KR^{2})\right\} \text{.}\label{eqnEpsilonMRelay}\end{equation}
 Notice that the effective spatial density is $\lambda\gamma/(m+1)$
and that $\epsilon$ is monotonically increasing with respect to $\lambda$,
we can immediately derive \begin{equation}
\frac{\lambda}{m+1}\leq~\frac{m\ln\frac{G{\pi}(1-\gamma)}{K{\gamma}}+{\ln}G-\ln(m+1)-\ln\ln\frac{1}{\epsilon}}{KR^{2}}\text{,}\end{equation}
 which yields (\ref{eqnTCMrelay}). Furthermore, in order to make
the capacity well-defined, i.e. $T_{m}(\epsilon)>0$, we will have
the constraint for outage probability stated in the corollary.

The derivation in a best-effort setting is exactly the same.

\begin{IEEEbiography}[{\includegraphics[width=1in,height=1.25in,clip,keepaspectratio]{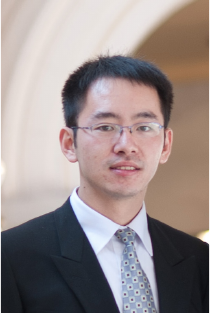}}]{Yuxin Chen}

received the B.S. in Microelectronics with High Distinction from Tsinghua
University in 2008, and the M.S. in Electrical and Computer Engineering
from the University of Texas at Austin. He is now a PhD student in
the Department of Electrical Engineering and a Master student in the
Department of Statistics at Stanford University. His research interests
include network science, information theory, and high-dimensional
statistics.

\end{IEEEbiography}

\begin{IEEEbiography}[{\includegraphics[width=1in,height=1.25in,clip,keepaspectratio]{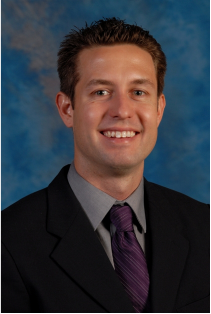}}]{Jeffrey G. Andrews}

(S\textquoteright{}98, M\textquoteright{}02, SM\textquoteright{}06)
received the B.S. in Engineering with High Distinction from Harvey
Mudd College in 1995, and the M.S. and Ph.D. in Electrical Engineering
from Stanford University in 1999 and 2002, respectively. He is a Professor
in the Department of Electrical and Computer Engineering at the University
of Texas at Austin, where he was the Director of the Wireless Networking
and Communications Group (WNCG) from 2008-12. He developed Code Division
Multiple Access systems at Qualcomm from 1995-97, and has consulted
for entities including the WiMAX Forum, Microsoft, Apple, Clearwire,
Palm, Sprint, ADC, and NASA. 

Dr. Andrews is co-author of two books, Fundamentals of WiMAX (Prentice-Hall,
2007) and Fundamentals of LTE (Prentice-Hall, 2010), and holds the
Earl and Margaret Brasfield Endowed Fellowship in Engineering at UT
Austin, where he received the ECE department\textquoteright{}s first
annual High Gain award for excellence in research. He is a Senior
Member of the IEEE, a Distinguished Lecturer for the IEEE Vehicular
Technology Society, served as an associate editor for the IEEE Transactions
on Wireless Communications from 2004-08, was the Chair of the 2010
IEEE Communication Theory Workshop, and is the Technical Program co-Chair
of ICC 2012 (Comm. Theory Symposium) and Globecom 2014. He has also
been a guest editor for two recent IEEE JSAC special issues on stochastic
geometry and femtocell networks.

Dr. Andrews received the National Science Foundation CAREER award
in 2007 and has been co-author of five best paper award recipients,
two at Globecom (2006 and 2009), Asilomar (2008), the 2010 IEEE Communications
Society Best Tutorial Paper Award, and the 2011 Communications Society
Heinrich Hertz Prize. His research interests are in communication
theory, information theory, and stochastic geometry applied to wireless
cellular and ad hoc networks.

\end{IEEEbiography}

\bibliographystyle{IEEEtran}

\bibliographystyle{IEEEtran} \bibliographystyle{IEEEtran}
\bibliography{bibfile}

\end{document}